\documentclass[]{aa}
\usepackage[latin1]{inputenc}
\usepackage[english]{babel}
\usepackage{amsmath}
\usepackage{amsfonts}
\usepackage{amssymb}
\usepackage[]{graphicx}
\usepackage{natbib,twoopt}
\usepackage[breaklinks=true]{hyperref} 
\bibpunct{(}{)}{;}{a}{}{,} 
\makeatletter
\newcommandtwoopt{\citeads}[3][][]{\href{http://adsabs.harvard.edu/abs/#3}%
{\def\hyper@linkstart##1##2{}%
\let\hyper@linkend\@empty\citealp[#1][#2]{#3}}}
\newcommandtwoopt{\citepads}[3][][]{\href{http://adsabs.harvard.edu/abs/#3}%
{\def\hyper@linkstart##1##2{}%
\let\hyper@linkend\@empty\citep[#1][#2]{#3}}}
\newcommandtwoopt{\citetads}[3][][]{\href{http://adsabs.harvard.edu/abs/#3}%
{\def\hyper@linkstart##1##2{}%
\let\hyper@linkend\@empty\citet[#1][#2]{#3}}}
\newcommandtwoopt{\citeyearads}[3][][]%
{\href{http://adsabs.harvard.edu/abs/#3}
{\def\hyper@linkstart##1##2{}%
\let\hyper@linkend\@empty\citeyear[#1][#2]{#3}}}
\makeatother

\title{Estimation and correction of wavefront aberrations using the self-coherent camera: laboratory results} 

\titlerunning{Estimation \& correction of wavefront aberrations using the SCC}
\authorrunning{J. Mazoyer et al.}

\author{J.~Mazoyer\inst{\ref{inst1}}\and P.~Baudoz\inst{\ref{inst1}}\and
R.~Galicher\inst{\ref{inst1}}\and M.~Mas\inst{\ref{inst2}}\and G.~Rousset\inst{\ref{inst1}}}

\institute{LESIA, Observatoire de Paris, CNRS, UPMC Paris 6 and Denis Diderot Paris 7, Meudon, France. \email{johan.mazoyer@obspm.fr}\label{inst1} 
\and 
Laboratoire d'Astrophysique de Marseille, CNRS, Aix-Marseille Univ., Marseille, France\label{inst2}
}


\abstract 
{Direct imaging of exoplanets requires very high contrast levels, which are obtained using coronagraphs. But residual quasi-static aberrations create speckles in the focal plane downstream of the coronagraph which mask the planet. This problem appears in ground-based instruments as well as in space-based telescopes.}
{An active correction of these wavefront errors using a deformable mirror upstream of the coronagraph is mandatory, but conventional adaptive optics are limited by differential path aberrations. Dedicated techniques have to be implemented to measure phase and amplitude errors directly in the science focal plane.}
{First, we propose a method for estimating phase and amplitude aberrations upstream of a coronagraph from the speckle complex field in the downstream focal plane. Then, we present the self-coherent camera, which uses the coherence of light to spatially encode the focal plane speckles and retrieve the associated complex field. This enable us to estimate and compensate in a closed loop for the aberrations upstream of the coronagraph. We conducted numerical simulations as well as laboratory tests using a four-quadrant phase mask and a 32x32 actuator deformable mirror.}
{We demonstrated in the laboratory our capability to achieve a stable closed loop and compensate for phase and amplitude quasi-static aberrations. We determined the best-suited parameter values to implement our technique. Contrasts better than $10^{-6}$ between 2 and 12 $\lambda/D$ and even $3.10^{-7}$ (RMS) between 7 and 11 $\lambda/D$ were reached in the focal plane. It seems that the contrast level is mainly limited by amplitude defects created by the surface of the deformable mirror and by the dynamic of the detector.} 
{These results are promising for a future application to a dedicated space mission for exoplanet characterization. A number of possible improvements have been identified.}

\begin{document}
\maketitle

\section{Introduction}
\label{sec:intro}  

Direct imaging is crucial to increase our knowledge of extra-solar planetary systems. On the one hand, it can detect long-orbit planets that are inaccessible for other methods (transits, radial velocities). On the other hand, it allows the full spectroscopic characterization of the surface and atmosphere of exoplanets. In a few favorable cases, direct imaging has already enabled the detection of exoplanets \citep{Kalas08,Lagrange_Bpic09} and even of planetary systems \citep{Marois08,Marois10}. However, the main difficulties of this method are the high contrast and small separation between the star and its planet. Indeed, a contrast level of $10^{-10}$ has to be reached within a separation of $\sim0.1''$ or lower to allow the detection of rocky planets.

To reduce the star light in the focal plane of a telescope, several coronagraphs have been developed, such as the four-quadrant phase mask (FQPM) coronograph \citep{Rouan00}, the vortex coronograph \citep{Mawet05} and the phase-induced amplitude apodization coronograph \citep{Guyon05}. However, the performance of these instruments is drastically limited by phase and amplitude errors. Indeed, these wavefront aberrations induce stellar speckles in the image, which are leaks of the star light in the focal plane downstream of the coronagraph. When classical adaptive optics (AO) systems correct for most of the dynamic wavefront errors that are caused by to atmosphere, they use a dedicated optical channel for the wavefront sensing. Thus, they cannot detect quasi-static non-common path aberrations (NCPA) created in the differential optical paths by the instrument optics themselves. These NCPA have to be compensated for using dedicated techniques, for ground-based telescopes as well as for space-based instruments.

Two strategies have been implemented to overcome the quasi-static speckle limitation. First, one can use differential imaging techniques to calibrate the speckle noise in the focal plane. Theses methods can use either the spectral signature and polarization state of the planet or differential rotation in the image \citep{Marois04,Marois06a}. Second, even before applying these post-processing techniques, an active suppression of speckles \citep{Malbet95} has to be implemented to reach very high contrasts. It uses a deformable mirror (DM) controlled by a specific wavefront sensor that is immune against NCPA. The techniques developed for this purpose include dedicated instrumental designs \citep{guyonAPJ09,Wallace10}, or creating of known phases on the DM \citep{BordeTraub06,Giveon07} to estimate the complex speckle field.

Ground-based instruments that combine these two strategies are currently being developed, such as SPHERE \citep{Beuzit08} and GPI \citep{Macintosh08}, to detect young Jupiter-like planets with an expected contrast performance of $10^{-6}$ at $0.5''$. Better contrasts might be achieved to reach the rocky planet level with instruments using dedicated active correction techniques embedded in space telescopes \citep{Trauger07}.

In this context, we study a technique of wavefront sensing in the focal plane that allows an active correction in a closed loop. This paper has two main objectives. First we give an overview of how the amplitude and phase errors upstream of a coronagraph can be retrieved from the complex amplitude of the speckle field (Section~\ref{sec:corono_estimation}) and how they can be compensated for using a DM (Section~\ref{sec:speck_corr}). In Section~\ref{sec:SCC_estim}, we introduce the self-coherent camera \citep{Baudoz06,Galicher08}. This instrument uses the coherence of the stellar light to generate Fizeau fringes in the focal plane and spatially encode the speckles. Using both the aberration estimator and the self-coherent camera (SCC), we are able to correct phase and amplitude aberrations. The second objective of the paper is a laboratory demonstration of the active correction and an experimental parametric study of the SCC (Section~\ref{sec:SCCCorrection}).

\section{Wavefront estimator in the focal plane of a coronagraph}
\label{sec:corono_estimation}  

In this section, we aim to prove that one can retrieve the wavefront upstream of the coronagraph using the measured complex amplitude of the electric field in the focal plane downstream of the coronagraph. We assume in the whole section that we can measure this complex amplitude without error using an undetermined method. We describe one type of this method (the SCC) in Section~\ref{sec:SCC_estim}. In Section~\ref{sec:coro_model}, we express the complex electric field that is associated to the speckles as a function of the wavefront errors in the pupil upstream of a phase mask coronagraph. From this expression, we propose an estimator of the wavefront errors from the speckle electric field (Section~\ref{sec:prop_estim}) and analyze its accuracy for an FQPM (Section~\ref{sec:acc_estim}).

\subsection{Expression of the complex amplitude of speckles in the focal plane as a function of the initial wavefront}
\label{sec:coro_model}

\begin{figure}[!ht]
   \begin{center}
   \includegraphics[width=9cm]{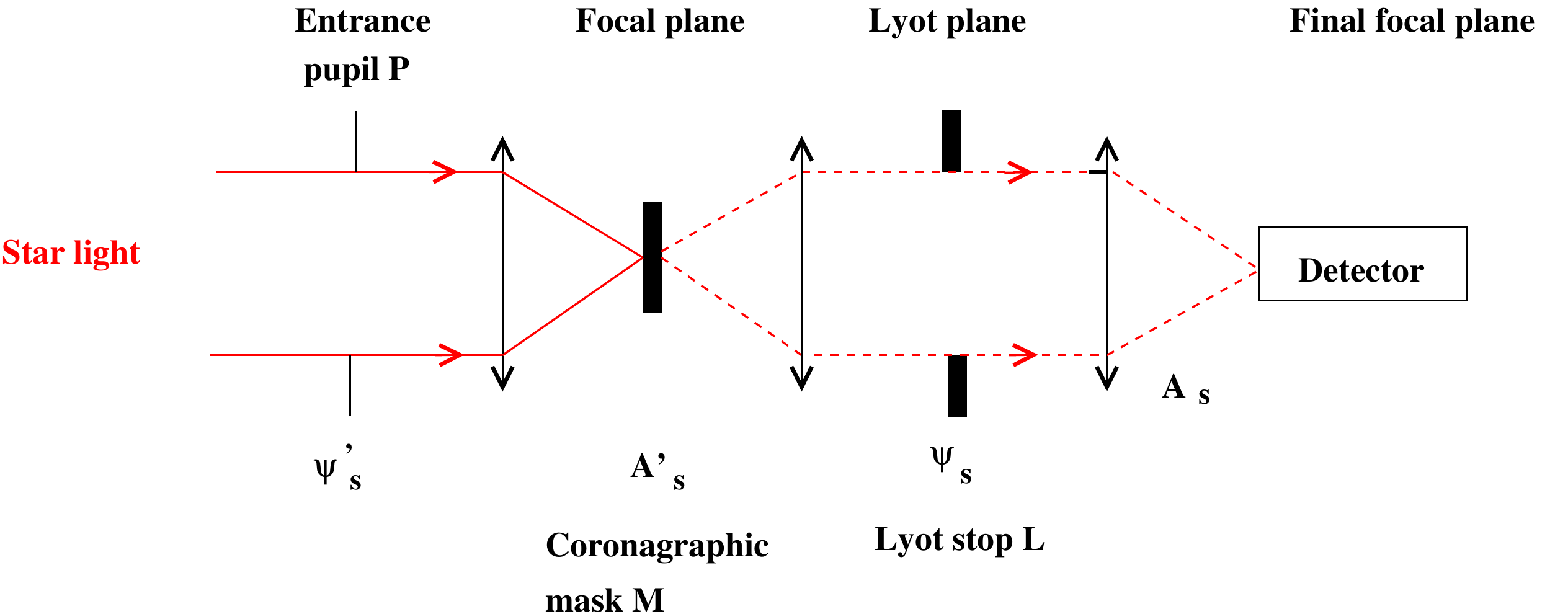}
   \includegraphics[height=2.5cm]{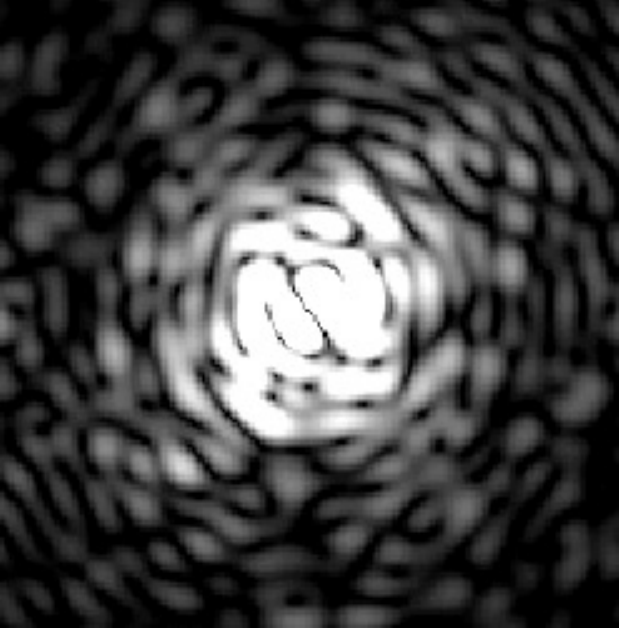}
   \end{center}
   \caption[example] 
   { \label{fig:speckles_sans_franges} 
\emph{Principle of a coronagraph (top). Aberrations in the entrance pupil plane induce speckles in the focal plane (bottom).}}
   \end{figure}

We consider here a model of a phase mask coronagraph using Fourier optics. Figure~\ref{fig:speckles_sans_franges} (top) presents the principle of a coronagraph. We assume that the star is a spatially unresolved monochromatic source centered on the optical axis. The stellar light moves through the entrance pupil~$P$. Behind this pupil, the beam is focused on the mask~$M$ in the focal plane, which diffracts the light. Hence, the non aberrated part of the stellar light is rejected outside of the imaged pupil in the next pupil plane and is stopped by the Lyot stop diaphragm~$L$. The aberrated part of the beam goes through the Lyot stop, producing speckles on the detector in the final focal plane (Figure~\ref{fig:speckles_sans_franges}, bottom).

We note whith~$\alpha$ and~$\phi$ the amplitude and phase aberrations in the entrance pupil plane and define the complex wavefront aberrations $\Phi$ as 
\begin{equation}
\Phi = \phi + i\alpha.
\end{equation} 
The complex amplitude of the star in the entrance pupil plane $\psi'_S$ can be written as
\begin{equation}
\psi'_S(\vec{\xi},\lambda) =\psi_0\,P(\vec{\xi})\,\exp \left(i\Phi(\vec{\xi})\right),
\end{equation} 
where $\psi_0$ is the mean amplitude of the field over the pupil P, $\vec{\xi}$ the coordinate in the pupil plane and $\lambda$ the wavelength. We assume that the aberrations are small and defined in the pupil~$P$ ($P\Phi = \Phi$), thus
\begin{equation}
\label{eq:psi_pr_s}
\dfrac{\psi'_S(\vec{\xi},\lambda)}{\psi_0} \simeq P(\vec{\xi})+i\Phi(\vec{\xi}).
\end{equation}
The complex amplitude of the electric field $A'_S$ behind the coronagraphic mask~$M$ in the first focal plane is
\begin{equation}
\label{eq:a_pr_s}
A'_S= \mathcal{F}[\psi'_S]\,M,
\end{equation}
 where $\mathcal{F}$ is the Fourier transform. Using Equation~\ref{eq:psi_pr_s}, we can write the electric field $\mathcal{F}^{-1}(A'_S)$ before the Lyot stop
\begin{equation}
\label{eq:ampl_before_Lyot}
\dfrac{\mathcal{F}^{-1}[A'_S]}{\psi_0} = P*\mathcal{F}^{-1}[M] + i\Phi*\mathcal{F}^{-1}[M],
\end{equation}
where~$*$ is the convolution product. We call $\Phi_M$ the aberrated part of the field after the coronagraph:
\begin{equation}
\label{eq:ampl_phiM}
\Phi_M= \phi_M + i\alpha_M = \Phi*\mathcal{F}^{-1}[M].
\end{equation}
After the Lyot stop~$L$, the electric field $\psi_S$ is
\begin{equation}
\label{eq:psi_s}
\dfrac{\psi_S}{\psi_0} = (P*\mathcal{F}^{-1}[M])\,.L + i\Phi_M\,.L.
\end{equation}
We assume a coronagraph for which the non aberrated part of the electric field is null inside the imaged pupil. This property of the perfect coronagraph~\citep{Cavarroc06} has also been demonstrated analytically for several phase coronagraphs such as FQPM coronagraphs~\citep{Abe03} and vortex coronagraphs~\citep{Mawet05}. The remaining part $\Phi_L$ of the normalized electric field after the Lyot stop reads
\begin{equation}
\label{eq:psi_Lyot}
\Phi_L = \phi_L + i\alpha_L = \Phi_M\,.L = (\Phi*\mathcal{F}^{-1}[M])\,.L.
\end{equation}
In the final focal plane, the complex amplitude $A_S$ is
\begin{equation}
\label{eq:ampl_focal_final}
\begin{array}{c}
A_S = \psi_0\mathcal{F}[i\Phi_L],\\
A_S = \psi_0\mathcal{F}[i(\Phi*\mathcal{F}^{-1}[M])\,.L],\\
A_S =  i\psi_0(\mathcal{F}[\Phi]\,.M)*\mathcal{F}[L].
 \end{array}
\end{equation}
This complex amplitude is directly related to the wavefront aberrations in the entrance pupil. If one can measure $A_S$, we can invert Equation~\ref{eq:ampl_focal_final} and retrieve the complex wavefront errors $\Phi$ in the entrance pupil.

\subsection{Wavefront estimator}
\label{sec:prop_estim}

We still assume in this section that an undefined method provides access to $A_S$. Using this complex amplitude $A_S$ as the measurement, we therefore propose the following estimator $\Phi_{est}$ for the wavefront:
\begin{equation}
\label{eq:estim_phi_ac_As}
\Phi_{est}= i\mathcal{F}^{-1}\left[\dfrac{A_S}{M\psi_0}\right].P.
\end{equation}
This estimator can be used for any phase mask coronograph (for which $M$ is nonzero over the full focal plane). To justify the pertinence of this estimator, we can re-write it using the variables of our model. Using Equation~\ref{eq:ampl_focal_final}, in a noise-free measurement case, this estimator reads
\begin{equation}
\label{eq:estim_phi_ss_As}
	  \Phi_{est}  = \left[((\Phi*\mathcal{F}^{-1}[M])\,.L)*\mathcal{F}^{-1}\left[\frac{1}{M}\right]\right]\,.P.
\end{equation}

Theoretically, if no Lyot stop is applied ($L=1$), Equation~\ref{eq:estim_phi_ss_As} becomes $\Phi_{est} = P\Phi = \Phi$. We propose this estimator based on the assumption that most of the information about the aberrations is not diffracted outside of the imaged pupil by the coronagraphic mask. Therefore, using this assumption, we intuit that for $L \simeq P$, we still have $\Phi_{est} \simeq \Phi$. This assumption is verified by the simulation in Section~\ref{sec:acc_estim}, and by the experiment described in Section~\ref{sec:SCCCorrection}.

\begin{figure}
   \begin{center}
   \includegraphics[width=9cm]{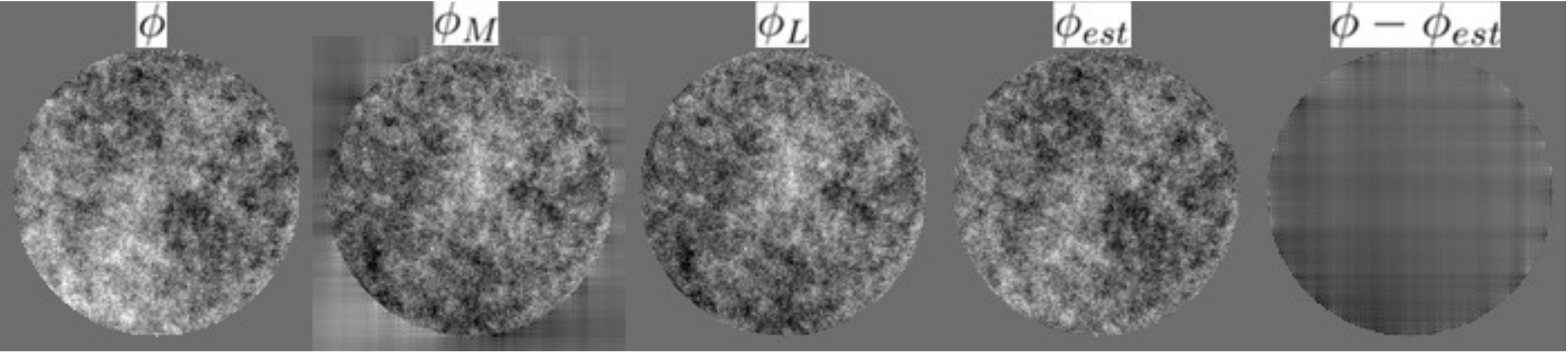}
   \end{center}
   \caption[example] 
   { \label{fig:Lyot_100_def} 
\emph{Simulations of an aberrated phase in the entrance pupil plane ($\phi$), and the real part of the field in the next pupil plane before ($\phi_M$) or after ($\phi_L$) the Lyot pupil. We also show the estimate ($\phi_{est}$) and the difference between $\phi_{est}$ and $\phi$. L = P in this case.}}
   \end{figure}

For a symmetrical phase mask such as the FQPM, either $\mathcal{F}^{-1}[M]$ and $\mathcal{F}^{-1}\left[\frac{1}{M}\right]$ are real. Thus, in the estimator we can separate the real ($\phi_{est}$) and imaginary part ($\alpha_{est}$) of the estimator in Equation~\ref{eq:estim_phi_ss_As}:
\begin{equation}
\label{eq:ph_est_TFreal}
   \left \{
   \begin{array}{c}
      \phi_{est} = \left[((\phi *\mathcal{F}^{-1}[M]).L)*\mathcal{F}^{-1}\left[\frac{1}{M}\right]\right].P \\
      \alpha_{est} = \left[((\alpha *\mathcal{F}^{-1}[M]).L)*\mathcal{F}^{-1}\left[\frac{1}{M}\right]\right].P.
   \end{array}
   \right .
\end{equation}
This relation ensures that within the limits of our model, this estimator independently provides estimates of the phase and amplitude aberrations.

\subsection{Performance of the estimation}
\label{sec:acc_estim}

In this section we test the accuracy of the estimation $\phi_{est}$ for a phase aberration $\phi$ and no amplitude aberrations ($\alpha = 0$). In the following numerical simulations, we assumed an FQPM coronagraph. It induces a phase shift of $\pi$ in two quadrants with respect to the two others quadrants. We simulated FQPM coronagraphs in this paper using the method described in \cite{Mas12}. This coronagraph is completely insensitive to some aberrations, for instance to one of the astigmatism aberrations \citep{Galicher_these,Galicher10}. Because these aberrations introduce no aberration inside the Lyot pupil, we are unable to estimate them. We assumed an initial phase with aberrations of 30 nm root mean square (RMS) over the pupil at $\lambda = 635nm$, with a power spectral density (PSD) in $f^{-2}$, where $f$ is the spatial frequency. 

In these simulations, we studied two cases. First, we used a Lyot pupil of the same diameter as the entrance pupil ($L=P$). Then, we studied the case of a reduced Lyot ($D_L < D_P$, where $D_L$ and $D_P$ are the diameters of the Lyot and entrance pupil, respectively).

\subsubsection{Case $L=P$}

Figure~\ref{fig:Lyot_100_def} shows the effect of phase-only aberrations $\phi$ in different planes of the coronagraph. Starting from the left, we represent the initial phase $\phi$, the real part of the amplitude due to the aberrations $\phi_M$ ($\phi_L$) before the Lyot stop (after the Lyot stop), derived from Equation~\ref{eq:ampl_phiM} (Equation~\ref{eq:psi_Lyot}) for phase-only aberrations. The last two images are the estimator $\phi_{est}$ and the difference between the estimate and the entrance phase aberrations ($\phi - \phi_{est}$). 

The estimate $\phi_{est}$ is very close to the initial phase $\phi$. For initial phase aberrations of $30$ nm RMS, the difference $\phi-\phi_{est}$ presents a level of 10 nm RMS in the entire pupil. The vertical and horizontal structures in this difference are due to the cut-off by the Lyot stop of the light diffracted by the FQPM (the light removed between $\phi_M$ and $\phi_L$), which leads to an imperfect estimate of the defects on the pupil edges. Aberrations to which the FQPM coronagraph is not sensitive (such as astigmatism) are also present in this difference.

Assuming a perfect DM, we can directly subtract $\phi_{est}$ from $\phi$ in the entrance pupil. Then, we can estimate the residual error once again, and iterate the process. The aberrations in the Lyot pupil $\phi_{L}$ converge toward zero (0.2 nm in ten iterations). This is important because these aberrations are directly linked to the speckle intensity in the focal plane downstream of the coronagraph. However, the difference $\phi- \phi_{est}$ does not converge toward zero in the entrance pupil. The fact that $\phi_{L}$ converges toward zero proves that the residual phase is only composed of aberrations unseen by the FQPM.

\subsubsection{Case $D_L < D_P$}

In a more realistic case, we aim to remove all the light diffracted by the coronagraphic mask, even for unavoidable misalignments of the Lyot stop. For this reason, the Lyot stop is often chosen to be slightly smaller than the imaged pupil. We consider here a Lyot stop pupil $L_{95\%}$ whose diameter is $D_L=95\% D_P$. In a first part, we show below that phase defects at the edge of the entrance pupil can be partially retrieved, then we study the convergence of the estimator in this case. 

\begin{figure}
   \begin{center}
	
   \includegraphics[width=9cm]{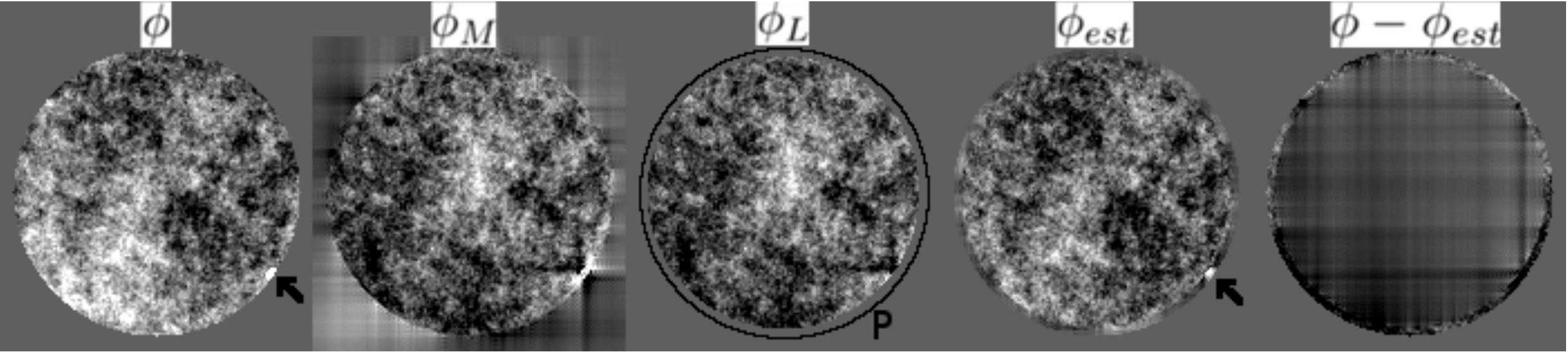}
   \end{center}
   \caption[example] 
   { \label{fig:Lyot_95_def} 
\emph{Simulations of an aberrated phase with a localized default in the entrance pupil plane ($\phi$), and in the next pupil plane ($\phi_M$, $\phi_L$). We represent the entrance pupil size by a dark ring around $\phi_L$. We show the estimate ($\phi_{est}$) and the difference between $\phi_{est}$ and $\phi$ in the last two images. $D_L=95\% D_P$ in this case.}}
   \end{figure}

As in Figure~\ref{fig:Lyot_100_def}, Figure~\ref{fig:Lyot_95_def} corresponds to the simulation of the consecutive steps of the model ($\phi$, $\phi_M$ before Lyot stops, $\phi_L$ after Lyot stop, then estimated phase $\phi_{est}$ and difference with initial phase). We added a small localized phase default, indicated by the black arrow, inside the entrance pupil~$P$, but outside of the Lyot stop~$L_{95\%}$ (Figure~\ref{fig:Lyot_95_def}, left). Around $\phi_L$, the complex amplitude after the Lyot stop, we drew a circle corresponding to the entrance pupil, slightly larger than $L_{95\%}$. For an FQPM, the additional defect is diffracted in the Lyot stop plane ($\phi_M$). After applying the Lyot stop of $95\%$ ($\phi_L$), most of  the default disappears, but we can still see its signature. As the estimator $\phi_{est}$ deconvolves by the phase mask, it partially retrieves the default, as seen in the estimate (indicated by the black arrow). In the error ($\phi - \phi_{est}$), we notice a remarkable cross issued from this defect, which is due to the information lost during the filtering by the Lyot stop. 

The wavefront estimation is limited when compared to the $D_L= D_P$ case. Because of the light filtered by the Lyot stop, some information about the wavefront aberrations close to the border of the entrance pupil is inevitably lost. Due to these unseen aberrations, $\phi-\phi_{est}$ does not converge toward zero. However, the residual aberrations in the Lyot pupil $\phi_{L}$ still converge toward zero, practically as quick as in the $L=P$ case.

By this first rough analysis, we see that we can estimate phase aberrations upstream of a coronagraph using the complex amplitude $A_S$ of the speckle field and Equation~\ref{eq:estim_phi_ac_As}. The same conclusion can be drawn for amplitude aberrations and, because the estimation is linear, for  a complex entrance wavefront. In the next section, we demonstrate that one can compensate for the wavefront errors in the entrance pupil.

\section{Entrance pupil wavefront correction}
\label{sec:speck_corr}

In this section, we use the estimator $\Phi_{est}$ to numerically simulate the correction of phase and amplitude aberrations in a closed loop. In the loop, we can remove constant factors in the estimator, which can be adjusted with a gain $g$
\begin{equation}
\label{eq:estim_As_and_gain}
\Phi_{est}= gi\mathcal{F}^{-1}\left[\dfrac{A_S}{M}\right].P.
\end{equation}
We still assume that we have a perfect sensor that measures the complex amplitude $A_S$ in the focal plane downstream of the coronagraph. We used a deformable mirror (DM) of NxN actuators upstream of the coronagraph, in the entrance pupil plane. We started with phase-only correction. We explain how to correct for the effects of phase and amplitude aberrations with only one DM in Section~\ref{sec:hermi_correc}.

We define the correction iterative loop by the expression of the residual phase $\phi^{j+1}$ at iteration $j+1$:
\begin{equation}
\label{eq:corDM}
	 \phi^{j+1}=\phi - \phi_{DM}^{j+1}=\phi - [\phi_{DM}^{j}+\sum_{i = 0}^{N^2 -1}k_i^{j+1} f_i],
\end{equation}
where $\phi_{DM}^{j}$ is the shape of the DM at iteration $j$, $k_i^{j+1}$ is the incremental command of the DM actuator $i$ at iteration $j+1$, and $f_i$ is the DM influence function, i.e., the WF deformation when only poking the actuator $i$. Note that $\phi^{j}$ is the phase to be estimated at iteration $j+1$.
The objective is now to determine the command vector $\{k_i^{j+1}\}$ from the phase estimator $\phi_{est}^{j+1}$. 

\subsection{Wavefront aberration minimization}
\label{sec:phase_minimize}

To derivate the command vector $\{k_i^{j+1}\}$, we minimize the distance between the measurements and the measurements that accounts for the parameters to be estimated. 
Ideally, we would like to find the $\lbrace k_i \rbrace$ minimizing the distance $d_{\lbrace k_i \rbrace}^j$ between the residual phase and the DM shape:
\begin{equation}
\label{eq:minimize0}
      d_{\lbrace k_i \rbrace}^j = \Vert  \phi^{j} - \sum_{i = 0}^{N^2 -1}k_i^{j+1} f_i \Vert^2.
\end{equation}
As presented in the previous section, a possible estimator of $\phi_j$ is given by Equation~\ref{eq:estim_As_and_gain}, allowing us to compute $\Phi_{est}^{j+1}$ from $A_S^{j}$ (directly related to $\phi^{j}$). So we minimize
\begin{equation}
\label{eq:minimize1}
       d_{\lbrace k_i \rbrace}^j = \Vert  \Re\left[ \Phi_{est}^{j+1}\right]  - D \{k_i^{j+1}\} \Vert^2,
\end{equation}
where $D$ is the interaction matrix because we are using a linear model. This matrix is calibrated off-line directly using the wavefront sensor \citep{Boyer90}. As in the conventional least-squares approach, we derived the pseudo inverse of $D$, denoted $D^\dag$, by the singular value decomposition (SVD) method. Therefore, the command vector solution of Equation~\ref{eq:minimize1} is given by
\begin{equation}
\label{eq:minimize_SVD}
   \{k_i^{j+1}\} = D^\dag \Re\left[ \Phi_{est}^{j+1}\right].
\end{equation}

\begin{figure}
   \begin{center}
   \begin{tabular}{c}
   \includegraphics[width=2.7cm, keepaspectratio=true]{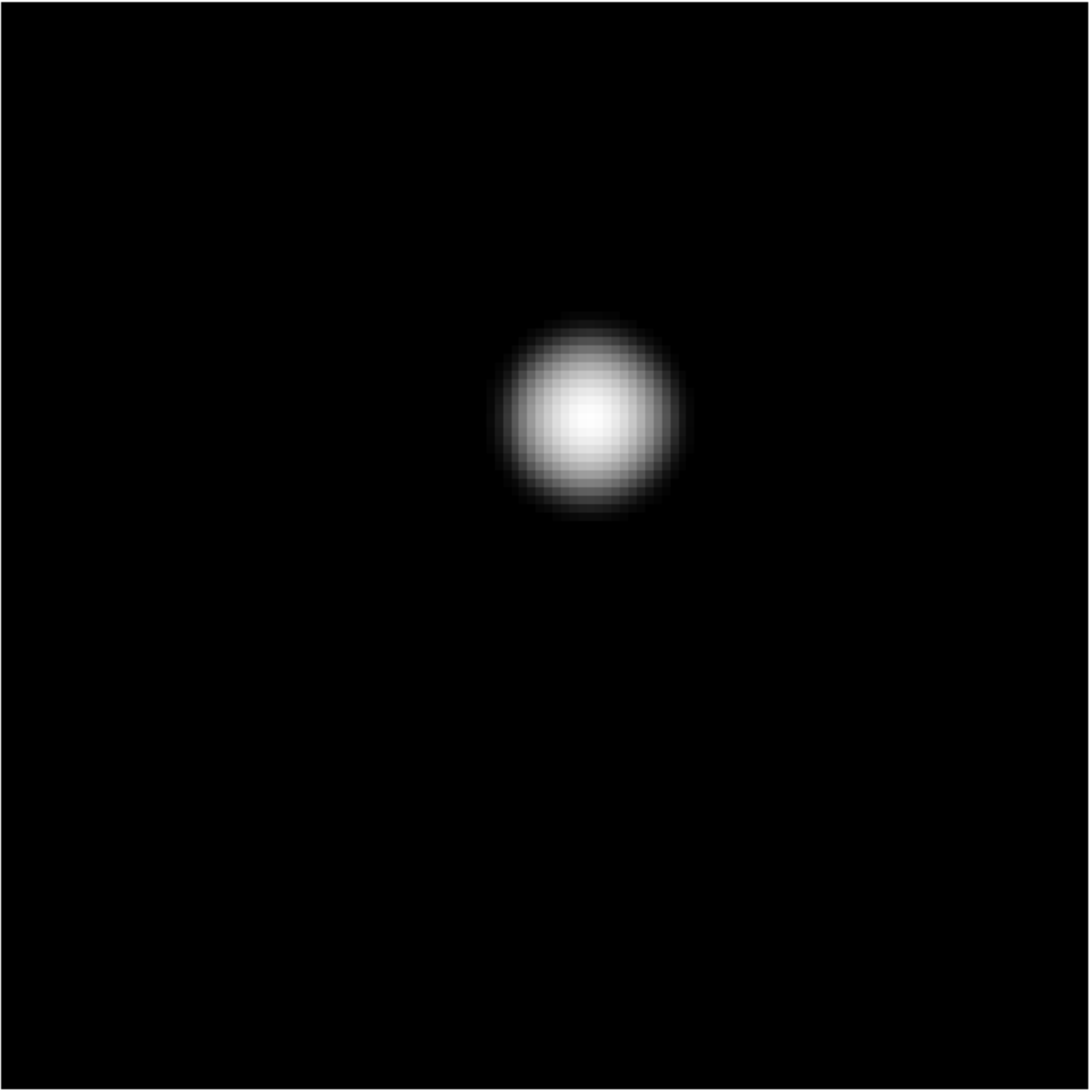}
   \includegraphics[width=2.7cm, keepaspectratio=true]{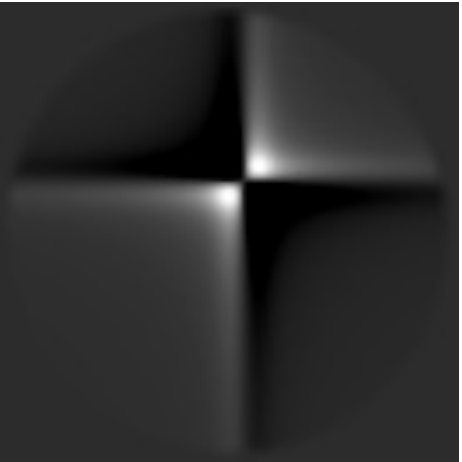}
   \includegraphics[width=2.7cm, keepaspectratio=true]{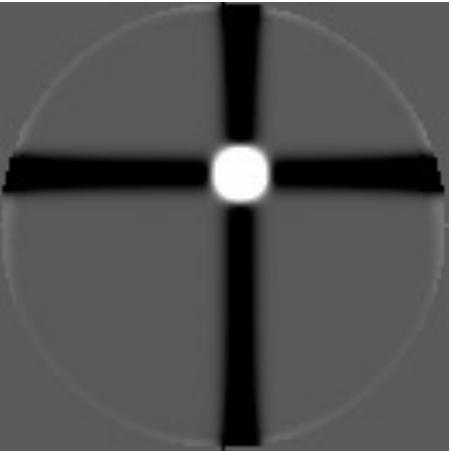}
   \end{tabular}
   \end{center}
   \caption[push_actionneur]  
   { \label{fig:push_actionneur} 
\emph{Simulations of the influence function $f_i$ in the pupil plane (left), and the effect using the two different estimators: we deconvolve by the mask ($\Phi_{est}$, right) or not ($\Phi_{est,2}$, center)}}
   \end{figure}

Equation~\ref{eq:minimize_SVD} is applicable for different estimators (only $D^\dag$ changes). To create the interaction matrix, we poked one by one the actuators while the others remain flat, as shown in Figure~\ref{fig:push_actionneur}. Each estimated phase vector obtained hence gives the column of the interaction matrix corresponding to the moved actuator. The influence function (which we simulated as a Gaussian function) is at the left, the estimator given by Equation~\ref{eq:estim_As_and_gain} at the right. At the center, we also plot another estimator of the wavefront that does not include the deconvolution by the coronagraph mask. It is defined by
\begin{equation}
\label{eq:minimize_model}
    \Phi_{est,2}= gi\mathcal{F}^{-1}\left[A_S\right].P.
\end{equation}
The chosen estimator applied to the influence function must be as spatially localized as possible: we have to filter the noise and it is far more efficient if the relevant information is gathered around one point. For this reason, it is preferable to use $\Phi_{est}$ (Figure~\ref{fig:push_actionneur}, right) instead of $\Phi_{est,2}$ (Figure~\ref{fig:push_actionneur}, center). We notice in Figure~\ref{fig:push_actionneur} that even after deconvolution by the FQPM, the estimate (right) shows a negative cross centered on the poked actuator, whereas it is not present in the initial phase (left). This artifact is generated to the transitions of the quadrants, which diffract the light outside of the Lyot stop. This cross can be a problem for two reasons. First, it is difficult to properly retrieve it in a noisy image. Then, because it enhances the cross-talk between the actuator estimates, it may lead to unstable corrections (see Section~\ref{sec:Correc_loop}).

For these reasons, we chose to create a synthetic interaction matrix, \emph{i.e.}, use a slightly different model for the estimation.

\subsection{Synthetic interaction matrix}
\label{sec:SyntMatr}  

One interest of interaction matrices is to calibrate the mis-registration between the DM and the wavefront sensor when considering a complex optical system. It also allows us to calibrate the shape and magnitude of each actuator response. Since the cross in Figure~\ref{fig:push_actionneur} (right) is 40 times less intense than the poke actuator in the center, it may be only partially retrieved in noisy images, which may lead to an unstable loop. To avoid this problem, we decided to build a synthetic interaction matrix based on the measured position and shape of each actuator. Because we considered an iterative measurement and correction loop, we finally discarded the magnitude calibration of each actuator, which lead to a slight increase of the required iteration number for convergence. We decided to only adjust the mis-registration and estimation shape of the actuator set in the output pupil on the measured interaction matrix, as seen by the sensor.  

Out of the NxN actuators of the square array in the DM, we chose to limit the response adjustment on the 12x12 actuators centered on the entrance pupil. These actuators were alternately pushed and pulled using known electrical voltages and we recorded $A_S$  for both positions. Assuming the complex amplitude $A_S$ is a linear function of the wavefront errors in the entrance pupil plane and that other aberrations of the optical path remain unchanged between two consecutive movements (the same pushed and pulled actuator), the difference between these two movements leads to the estimate $\hat{f_i}$ of the influence function of a single poked actuator using Equation~\ref{eq:estim_As_and_gain}. For each estimate, we adjusted a Gaussian function defined by its width and position in the output pupil. From these 144 Gaussian fits, we can build the actuator grid as observed in the plane, where aberrations are estimated and determine the inter-actuator distance in each direction and the orientation of this grid. We also determined the median width of the adjusted Gaussian functions and computed a synthetic Gaussian function, which was translated onto the adjusted actuator grid to create a new set of NxN synthetic estimates $\hat{f_i}^{synth}$.  

From these synthetic estimates, we built the synthetic interaction matrix $D^{synth}$. Some of the actuators are outside of the entrance pupil, and their impact inside the pupil is negligible. We excluded these actuators from $D^{synth}$. For any wavefront estimate, the distance to be minimized is now
\begin{equation}
\label{eq:minimize3}
	d_{\lbrace k_i \rbrace}^j = \Vert  \Re\left[ \Phi_{est}^{j+1}\right]  - D^{synth} \{k_i^{j+1}\} \Vert^2.
\end{equation}
The solution is given by the pseudo-inverse of the interaction matrix $D^{synth}$ using the SVD method.

\subsection{Phase and amplitude correction using one DM}
\label{sec:hermi_correc}

As explained in~\cite{BordeTraub06}, a complex wavefront $\Phi = \phi + i\alpha$ can be corrected for on half of the focal plane with only one DM. The idea is to apply a real phase on the DM to correct for the phase and amplitude on half of the focal plane. Because the Fourier transform of a Hermitian function is real, we define $A_S^{hermi}$ as 
\begin{equation}
\label{eq:hermi_def}
   \left \{
   \begin{array}{r c}
      \forall \vec{x} \in \mathbb{R} \times \mathbb{R}+ , &   A_S^{hermi}(\vec{x}) = A_S(\vec{x})\\
      \forall \vec{x} \in \mathbb{R} \times \mathbb{R}- ,  & A_S^{hermi}(\vec{x}) = A_S^*(-\vec{x})\\,
   \end{array}
   \right .
\end{equation}
where $A_S^*$ is the complex conjugated of $A_S$ and introduce it into the estimator of Equation~\ref{eq:estim_As_and_gain}. The resulting estimated wavefront is real, which allows its correction with only one DM.

Now we have a solution to correct for the wavefront aberrations upstream of the coronagraph when the complex amplitude in focal plane is known. We introduce in Section~\ref{sec:SCC_estim} a technique to measure $A_S$: the self-coherent camera.

\section{Self-coherent camera: a complex amplitude sensor in focal plane}
\label{sec:SCC_estim}

The self-coherent camera (SCC) is an instrument that allows complex electric field estimations in the focal plane.

\subsection{SCC principle}
\label{sec:SCCmodel}

\begin{figure}[!ht]
   \begin{center}

   \includegraphics[width=9cm]{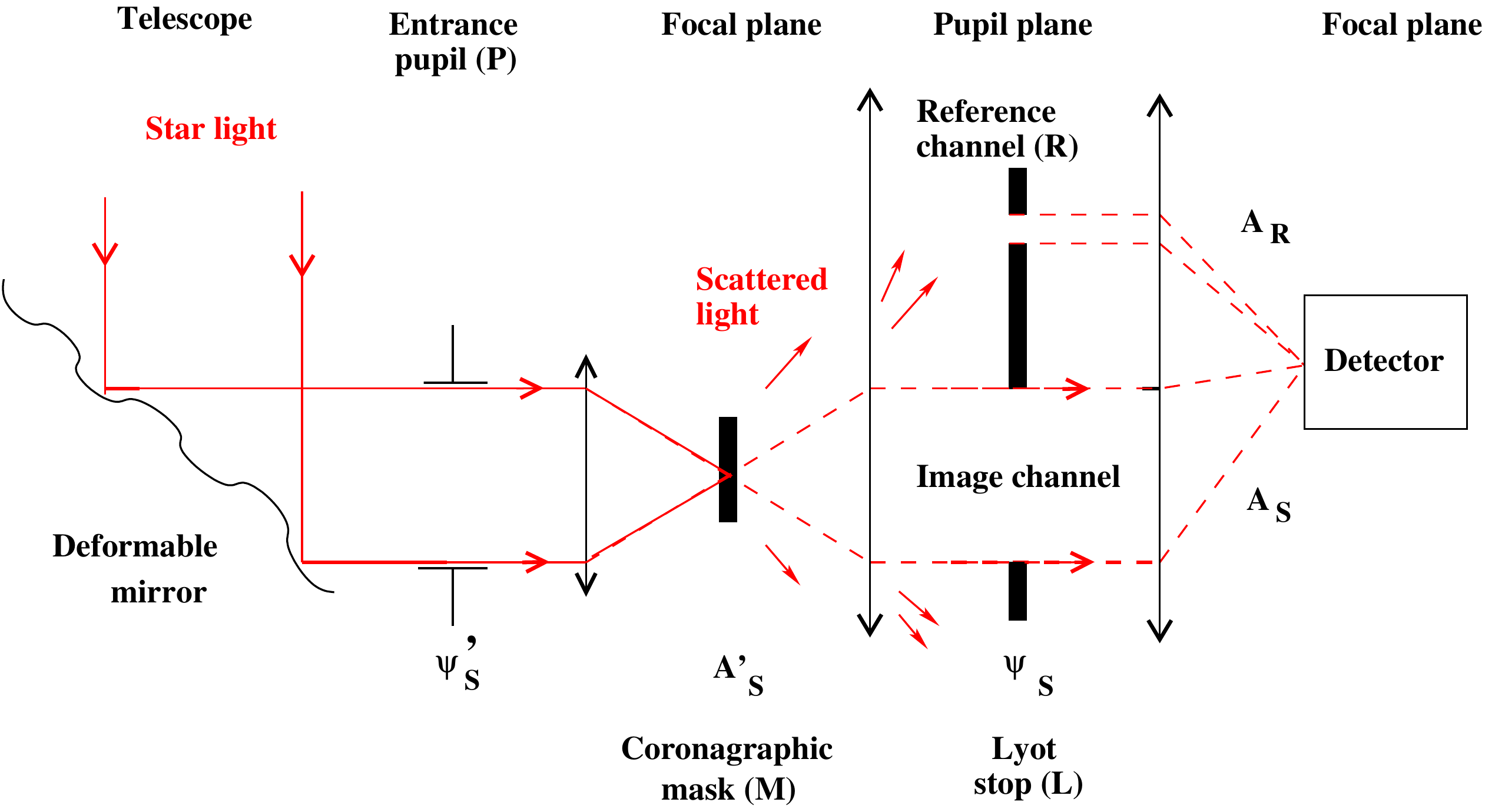}
   \includegraphics[height=2.5cm]{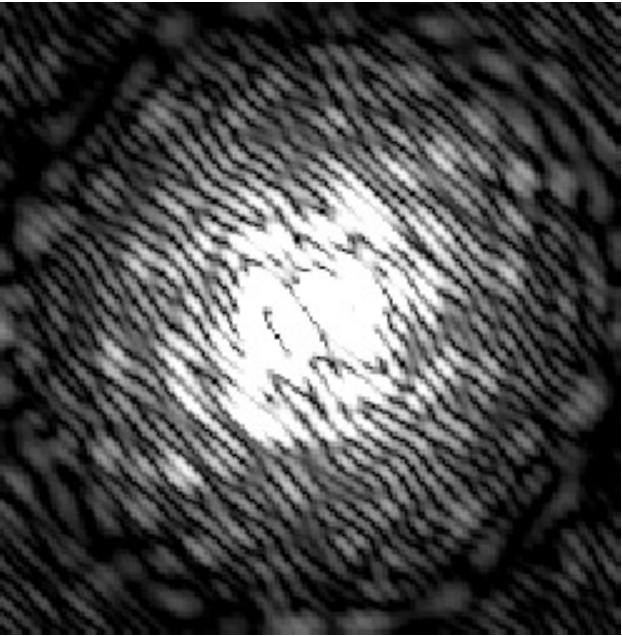}
   \end{center}
   \caption[example] 
   { \label{fig:SCC_principle} 
\emph{Principle of the SCC combined with a coronagraph and a DM (top). A small hole is added in the Lyot stop plane to create a reference channel. In the final focal plane (bottom), the SCC image is formed by speckles encoded with Fizeau fringes}}
   \end{figure}

Figure~\ref{fig:SCC_principle} (top) is a schematic representation of the SCC combined with a focal phase mask coronagraph and a DM. We added a small pupil~$R$, called reference pupil, in the Lyot stop plane of a classical coronagraph.~$R$ selects part of the stellar light that is diffracted by the focal coronagraphic mask. The two beams are recombined in the focal plane, forming Fizeau fringes, which spatially modulate the speckles. In the following, we call the SCC image the image of the encoded speckles (Figure~\ref{fig:SCC_principle}, bottom). In this section, we briefly demonstrate that this spatial modulation allows us to retrieve the complex amplitude $A_S$. A more complete description of the instrument can be found in~\cite{Baudoz06},~\cite{Galicher08,Galicher10}.

The electric field $\psi$ in the pupil plane (Equation \ref{eq:psi_s}) after the modified Lyot stop is
\begin{equation}
\label{eq:model_SCC1}
   \begin{array}{r}
		\dfrac{\psi(\vec{\xi},\lambda)}{\psi_0} = \left[(P(\vec{\xi})+i\Phi(\vec{\xi},\lambda))*\mathcal{F}^{-1}[M](\vec{\xi})\right].\\
		\left(L(\vec{\xi})+R(\vec{\xi})*\delta(\vec{\xi}-\vec{\xi_0})\right),
   \end{array}
\end{equation}
where $\vec{\xi_0}$ is the separation between the two pupils in the Lyot stop, and $\delta$ is the Kronecker delta. $\psi$ can also be written as
\begin{equation}
\label{eq:model_SCC2}
	    \psi(\vec{\xi},\lambda) = \psi_S(\vec{\xi},\lambda)+\psi_R(\vec{\xi},\lambda)*\delta \left(\vec{\xi}-\vec{\xi_0}\right),
\end{equation}
where $\psi_S$ is the complex amplitude in the Lyot stop, defined in Equation~\ref{eq:psi_Lyot}, and $\psi_R$ is the complex amplitude in the reference pupil. We denote with $A_R$ its Fourier transform, the complex amplitude in the focal plane, of the light issued from the reference pupil. In monochromatic light, the intensity $I = \left|\mathcal{F}[\psi]\right|^2$ recorded on the detector in the final focal plane can then be written as
\begin{equation}
\label{eq:focal_plane}
   \begin{array}{c}
I(\vec{x}) = |A_S(\vec{x})|^2+|A_R(\vec{x})|^2\\
+A^*_S(\vec{x})\,A_R(\vec{x}) \exp\left(\frac{-2i\pi \vec{x}.\vec{\xi_0}}{\lambda}\right)\\
+A_S(\vec{x})\,A^*_R(\vec{x}) \exp\left(\frac{2i\pi \vec{x}.\vec{\xi_0}}{\lambda}\right),
   \end{array}
\end{equation}
where $A^*$ is the conjugate of $A$ and $\vec{x}$ the coordinate in the focal plane. The two first terms are the intensities issued from Lyot and reference pupils, and provide access only to the square modulus of the complex amplitudes. The two correlation terms that create the fringes directly depend on $A_S$  and $A_R$.

When an off-axis source (planet) is in the field of view, its light is not diffracted by the coronagraphic mask. Thus, it does not go through the reference pupil. Because the lights of the off-axis and in-axis sources are not coherent, the off-axis light amplitude in the focal plane does not appear in the correlation terms (\emph{i.e.}, its image is not fringed).

\subsection{Complex amplitude of the speckle field}
\label{sec:data_exctract}

In this section, we demonstrate that we can use the SCC image to estimate the complex amplitude of the speckle field. We first apply a numerical inverse Fourier transform to the recorded SCC image (Equation~\ref{eq:focal_plane}),
\begin{equation}
\label{eq:TF_I}
\begin{array}{c}
\mathcal{F}^{-1}[I](\vec{u}) = \mathcal{F}^{-1}[I_S+I_R]+\mathcal{F}^{-1}[A^*_S\,A_R]*\delta\left(\vec{u}-\frac{\vec{\xi_0}}{\lambda}\right)\\
+\mathcal{F}^{-1}[A_S\,A^*_R]*\delta\left(\vec{u}+\frac{\vec{\xi_0}}{\lambda}\right),
  \end{array}
\end{equation}
where $I_S = |A_S|^2$ and $I_R = |A_R|^2$ are the intensities of the speckles and reference pupil, and $\vec{u}$ is the coordinate in the Fourier plane.

\begin{figure}[!ht]
   \begin{center}
   \includegraphics[height=6cm]{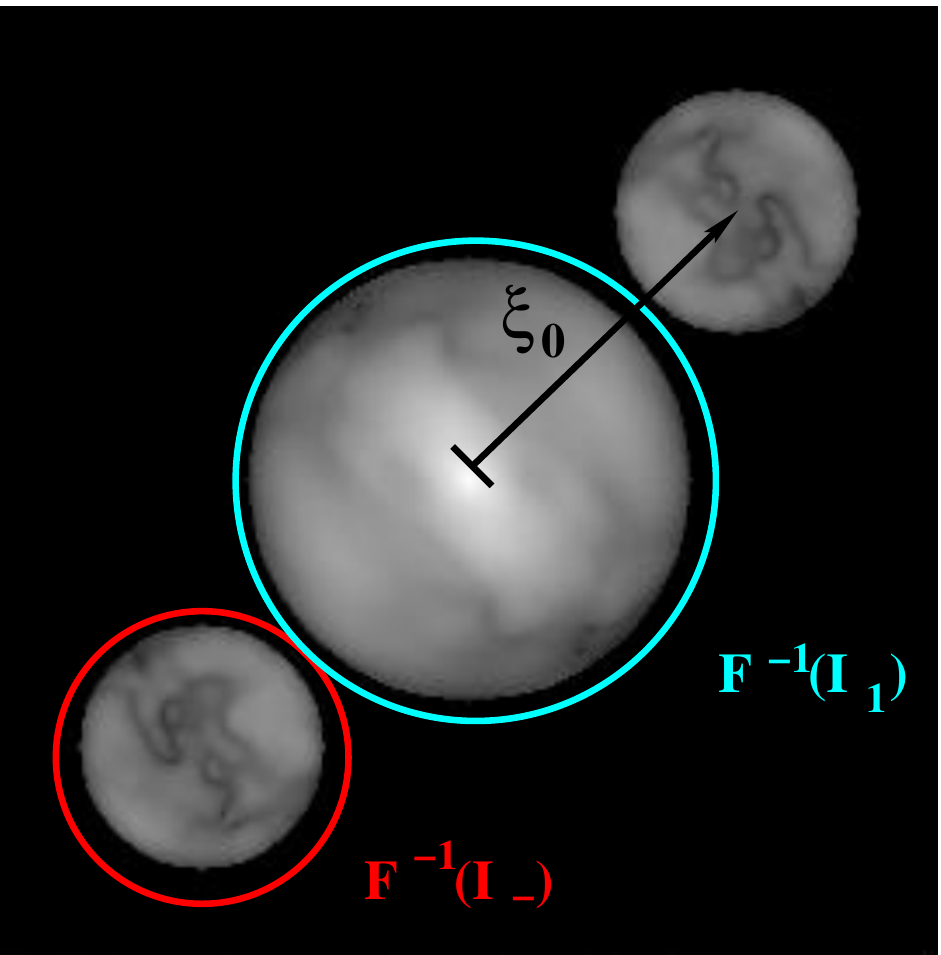}
   \end{center}
   \caption[example] 
   { \label{fig:3_pic} 
\emph{Correlation peaks in the Fourier transform of the focal plane. The inverse Fourier transform of $I_S + I_R$ is circled in blue. The inverse Fourier transform of $I_- = A_S\,A^*_R$ is circled in red.}}
   \end{figure} 
   
$\mathcal{F}^{-1}[I]$ is composed of three peaks centered at $\vec{u} = [-\vec{\xi_0}/\lambda, \vec{0}, +\vec{\xi_0}/\lambda]$ (Figure~\ref{fig:3_pic}). We denote with $D_L$ the diameter of the Lyot pupil and with $D_L/\gamma$ the diameter of the reference pupil ($\gamma >1$). The central peak is the sum of the autocorrelation of the Lyot and reference pupils $\mathcal{F}^{-1}[I_S+I_R]$. Its radius is $D_L$ because we assume $\gamma > 1$. The lateral peaks of the correlation ($\mathcal{F}^{-1}[I_-]$ and $\mathcal{F}^{-1}[I_+]$ hereafter) have a radius $(D_L+D_L/\gamma)/2$. Thus the three peaks do not overlap only if~\citep{Galicher10}
\begin{equation}
||\vec{\xi_0}|| > \frac{D_L}{2}\left(3+\frac{1}{\gamma} \right),
\end{equation}
which puts a condition on the smallest pupil separation. The lateral peaks are conjugated and contain information only on the complex amplitude of the stellar speckles that are spatially modulated on the detector. When we shift one of these lateral peaks to the center of the correlation plane ($\vec{u}=\vec{0}$), its expression can be derived from Equation \ref{eq:TF_I}: 
\begin{equation}
\label{eq:I_-}
\mathcal{F}^{-1}[I_-] = \mathcal{F}^{-1}[A_S\,A^*_R].
\end{equation}

Assuming $\gamma \gg 1$, we can consider that the complex amplitude in the reference pupil is uniform and that $A^*_R$ is the complex amplitude of an Airy pattern. Therefore, knowing $A_R$, we can to retrieve the complex amplitude $A_S$ in the focal plane using the SCC (where $A^*_R$ is not zero):
\begin{equation}
\label{eq:AS_function_I}
A_S = \frac{I_-}{A^*_R}.
\end{equation}

\subsection{SCC wavefront estimation}
\label{sec:SCCphase_estim}
   
Equation~\ref{eq:estim_As_and_gain} shows how to estimate the wavefront upstream of a coronagraph is estimated using the complex amplitude of the speckle field in the focal plane. Combining Equations~\ref{eq:estim_phi_ac_As}~and~\ref{eq:AS_function_I}, we have an estimator of the wavefront aberrations $\Phi$ as a function of $I_-$:
\begin{equation}
\label{eq:estimator}
\Phi_{est}= \left[i\mathcal{F}^{-1}\left[\dfrac{I_-}{A^*_R\psi_0 M}\right]\right]\,.P. \\
\end{equation}
This estimator is only limited in frequency by the size of the reference pupil. Indeed, where the reference flux is null, the speckles are not fringed and their estimate cannot be achieved. Small reference pupils produce large point spread functions (\emph{i.e.}, with a first dark ring at large separation) and allow estimating $A_S$ in a large area of the focal plane. The influence of the reference pupil size is detailed in Section~\ref{sec:SizeRef}. 

\begin{figure}
   \begin{center}
   \includegraphics[width=9cm]{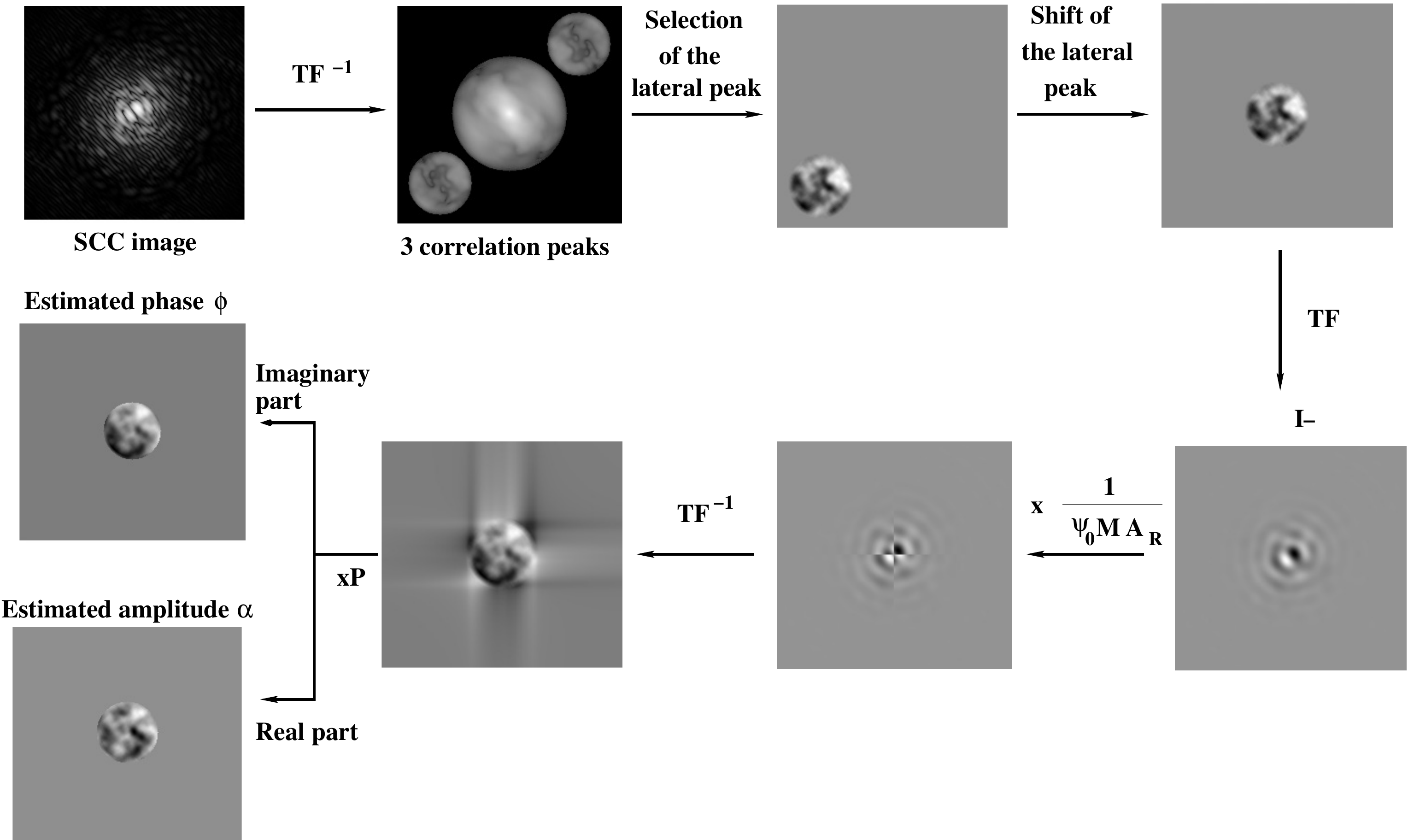}
   \end{center}
   \caption[example] 
   { \label{fig:step_estim_fo} 
\emph{Steps followed to estimate the phase and amplitude from SCC images.}}
   \end{figure}

Figure~\ref{fig:step_estim_fo} summarizes the steps followed to estimate the phase and amplitude aberrations with the SCC. From the fringed focal plane, we used a Fourier transform to retrieve $I_-$, from which we deduced the complex amplitude of the speckle field $A_S$. Using the estimator, we measured the phase and amplitude aberrations.

\subsection{Correction loop}
\label{sec:Correc_loop}

We can use this wavefront estimator to control a DM and correct for the speckle field in the focal plane as explained in Section~\ref{sec:speck_corr}. The DM has a finite number of degrees of freedom and thus can only correct for the focal plane in a limited zone. If the reference pupil is small enough ($\gamma \gg 1$), the point spread function (PSF) $|A_R|^2$ is uniform over the correction zone ($A^*_R \simeq A_0$). We discuss this assumption in Section~\ref{sec:SizeRef}. Under this assumption, Equation~\ref{eq:estimator} becomes
\begin{equation}
\label{eq:corr_operator_ac_SCC}
\Phi_{est} \simeq i\mathcal{F}^{-1}\left[\frac{I_-}{A_0\psi_0 M }\right]\,.P = gi\mathcal{F}^{-1}\left[\frac{I_-}{M}\right]\,.P.
\end{equation}
As described in Section~\ref{sec:speck_corr}, we removed the constant terms in the estimation and put them into the gain g. From $\Phi_{est}$, we created a synthetic matrix, as explained in Section~\ref{sec:SyntMatr}. Similarly, the other estimator $\Phi_{est,2} $ introduced in Equation~\ref{eq:minimize_model}, becomes
\begin{equation}
\label{eq:corr_operator2_ac_SCC_ac_gain}
\Phi_{est,2} = gi\mathcal{F}^{-1}\left[I_-\right]\,.P.
\end{equation}
Using the interaction matrices deduced from these estimators ($\Phi_{est}$, $\Phi_{est,2}$) and the synthetic one, we studied the correction loop. We simulated a DM with 27 actuators across the entrance pupil. To build these matrices, we only selected the actuators with a high influence in the pupil (633 actuators were selected for this number of actuators in the pupil). Lyot stop and entrance pupil have the same radius, and we chose $\gamma = 40$ for the reference pupil size.

\begin{figure}
   \begin{center}
   \includegraphics[width=8cm]{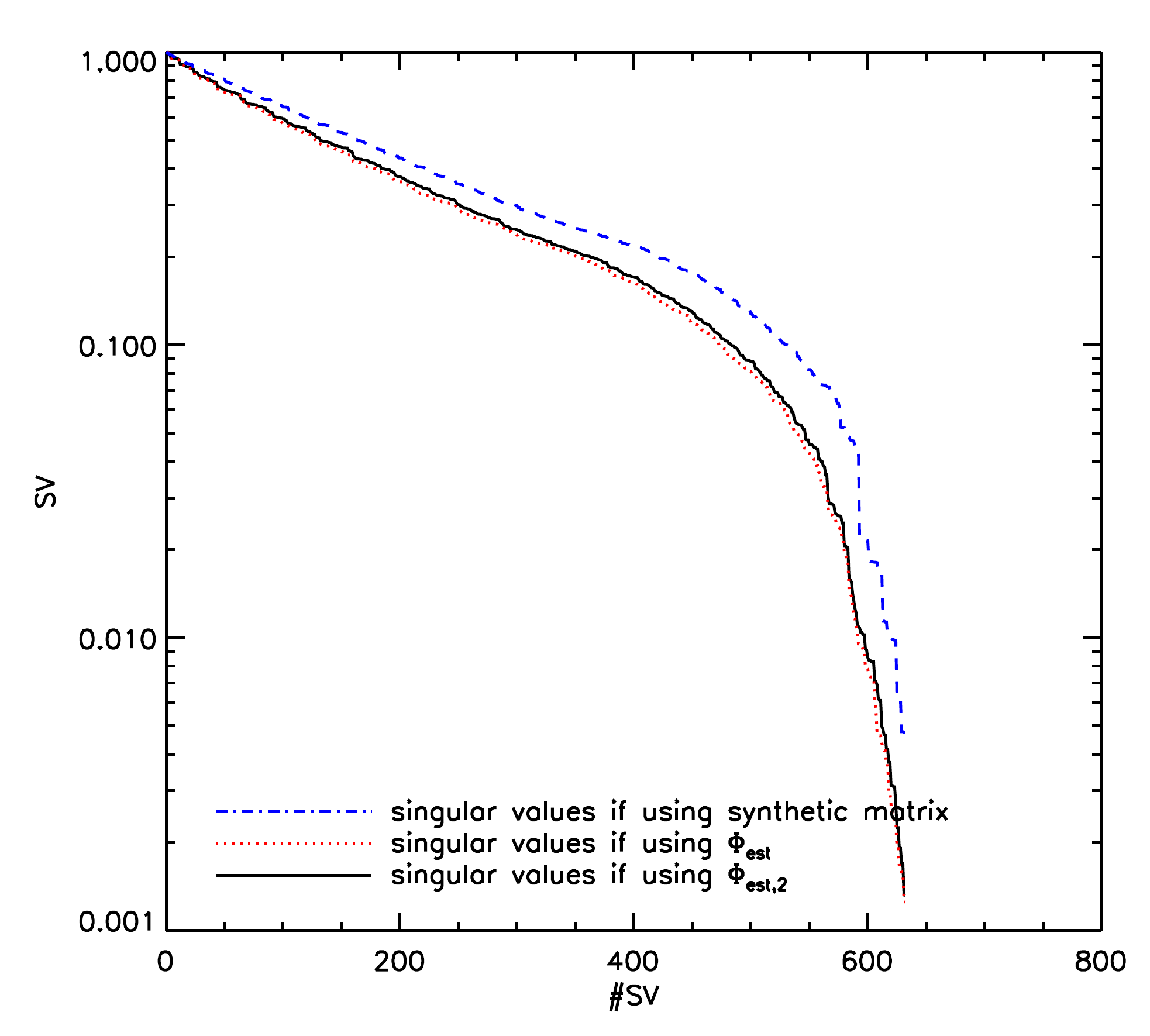} 
   \end{center}
   \caption[val_sing] 
   { \label{fig:val_sing} 
\emph{Singular values, normalized to their respective maximum, issued from the inversion of the interaction matrices $D$, obtained using the two estimators $\Phi_{est}$ (red,dotted) and $\Phi_{est,2}$ (black, solid) and the synthetic matrix (blue, dashed) for $\gamma = 40$.}}
   \end{figure}

In Figure~\ref{fig:val_sing}, we plot the singular values (SV), normalized to their highest values, derived from the inversion of the matrices $D$ obtained using the estimators $\Phi_{est}$ and $\Phi_{est,2}$ and of the interaction matrix built from $\hat{f_i}^{synth} $. As already underlined, the cross in $\Phi_{est}$ or $\Phi_{est,2}$ (Figure~\ref{fig:push_actionneur}, center and right) correlates the estimates of different actuators and therefore leads to lower SV (up to five times lower for the lowest SV). When inverted in $D^\dag$, low SV lead to higher values (in absolute values) and amplify the noise in Equation~\ref{eq:minimize_SVD}. Applied to noisy data, such $D^\dag$ matrices may lead to an unstable correction. Even in a noise-free case, simulations of the correction with the three methods and the same number of actuators used (633) showed that only the synthetic matrix leads to a stable correction.

\subsection{Optical path difference}
\label{sec:OPD}

Between the Lyot stop and the detector, the beam is split into two paths (image and reference), which encounter different areas in the optics. Thus, differential aberrations exist~\cite{Galicher10}. However, because the reference pupil is small ($\gamma \gg 1$), the main aberration is an optical path difference (OPD) between the two channels. In this section we study how this OPD impacts the SCC performance.

\subsubsection{Influence of an OPD on the correction}
\label{sec:OPD_infl}

Given an OPD $d_{op}$, we can define phase difference $\phi_{op} = 2\pi d_{op}/\lambda$. This phase difference modifies the $I_-$ originally defined in Equation~\ref{eq:I_-}:
\begin{equation}
	\label{eq:errro_I_-_ddm}
\ I_- = A_S\,A^*_R\exp(i\phi_{op}).
	\end{equation}
The phase and amplitude estimate $\phi_{est,op}$ and $\alpha_{est,op}$ can be expressed as a function of the estimates made without an OPD ($d_{op} = 0$):
\begin{equation}
	\label{eq:errro_ddm}
\ \phi_{est,op} + i\alpha_{est,op}  = (\phi_{est} + i\alpha_{est})\exp(i\phi_{op}),
	\end{equation}
and thus
\begin{equation}
	\label{eq:errro_ddm2}
\  \left \{
\ \begin{array}{r}
\  \phi_{est,op} =  \phi_{est}\cos(\phi_{op}) - \alpha_{est}\sin(\phi_{op})  \\
\  \alpha_{est,op} =  \phi_{est}\sin(\phi_{op}) +\alpha_{est}\cos(\phi_{op}).
\ \end{array}
 \  \right .
\end{equation}

Hence, even phase-only aberrations (such as the movements of the DM) have an influence on the estimated amplitude (\emph{i.e.}, the imaginary part of the estimator $\Phi_{est}$) for a nonzero OPD. In this section, we make two assumptions. First, that the DM is perfect and we can correct for any desired phase in the pupil plane. Second, that the only error in the estimator is due to the OPD: if $d_{op}=0$, the estimator retrieves the exact phase and amplitude ($\Phi_{est} = \phi + i\alpha$). 

We started the loop with a phase $\phi^{0}$ and an amplitude $\alpha$. After $j$ iterations the phase in the pupil plane $\phi^{j}$ is the difference between the previous phase $\phi^{j-1}$, and the estimate of this previous phase $\phi^{j-1}_{est,op}$. Under the previous assumptions, we have $\alpha_{est} = \alpha$ and $\phi_{est} = \phi^{j-1}$, and 
\begin{equation}
	\label{eq:errro_niter}
\ \phi^{j}  =\phi^{j-1} - \phi^{j-1}_{est,op} = \phi^{j-1}(1 - \cos(\phi_{op})) + \alpha\sin(\phi_{op}).
	\end{equation}
Because the OPD biases the estimation, the correction introduces an error at each iteration. This sequence converges if $|1 - \cos(\phi_{op})| < 1$. This assumption ($-\pi/2< \phi_{op} < \pi/2$) is always satisfied in real cases. Its limit $\phi_{end} $ satisfies the equation
\begin{equation}
\ \begin{array}{l}
\ \phi_{end}  = \phi_{end} (1 - \cos(\phi_{op})) + \alpha\sin(\phi_{op}) \\
\ \phi_{end} =\alpha\tan(\phi_{op}).
\ \end{array}
	\end{equation}
Therefore, for a nonzero OPD and a phase-only correction, the SCC correction converges, but the errors on the final phase depend on the uncorrected amplitude aberrations $\alpha$. 

To estimate the OPD effect on the level of the focal plane intensity, we considered the complex amplitude in the focal plane as a linear function of the phase and amplitude aberrations in the entrance pupil plane. We can thus evaluate the energy in the focal plane as a linear function of $|\phi|^2+|\alpha|^2$. Without an OPD, a perfect phase-only correction would leave a level of speckles only dependent on the entrance amplitude aberrations $|\alpha|^2$. With an OPD, this level is slightly higher: $|\alpha|^2(1+\tan(\phi_{op})^2)$. For a realistic phase difference of $0.1$ radians, the difference in intensity in the speckle field between the case with and without an OPD would be $1\%$. The impact on the correction is only weak.

The problem occurs when we try to correct phase and amplitude at the same time with one DM. We explained how to do this in Section~\ref{sec:hermi_correc}. For $\phi_{op} \neq 0$, numerical simulations as well as tests on an optical bench show that the correction is unstable: at each iteration, we raised the phase aberrations by trying to correct for the amplitude aberrations and \emph{vice versa}. Thus, we need an estimate of the OPD to stabilize the correction.

\subsubsection{Estimation and correction of the OPD}
\label{sec:OPDcorrec}

In the construction of the synthetic matrix, (Section~\ref{sec:SyntMatr}), we studied the difference of two SCC images produced by wavefronts that only differ by a movement of an actuator. Because the DM is in the pupil plane, the estimator applied to this difference is real for of an OPD equal to zero. However, for a nonzero OPD $d_{op}$ and using Equation~\ref{eq:errro_I_-_ddm} with $\alpha = 0$, we deduce $\Phi_{est,op}^i= \Phi_{est}^i (\sin(d_{op})+i\cos(d_{op}))$. For each of the 12x12 actuators used to build the synthetic matrix, the arctangent of the ratio of the imaginary part on the real part of $\Phi_{est,op}^i$ leads to an estimate of the OPD. Due to the noise in the image, small differences in the OPD estimate can appear from one actuator to another. Calculating the median of the estimated OPDs, we obtain the measured phase difference $\phi_{op}^{mes}$. We modified $I_-$ accounting for this OPD and our estimator (Equation~\ref{eq:corr_operator_ac_SCC}) becomes
\begin{equation}
	\label{eq:corr_operator_ac_SCC_opd}
	\Phi_{est,op} =gi\mathcal{F}^{-1}\left[\frac{I_-\exp(-i\phi_{op}^{mes})}{M}\right]\,.P.
\end{equation}
We use this new estimator from now on.

The OPD variations during the correction are a problem that has to be carefully considered for a telescope application. In the current installation (bench under a hood, room temperature stabilized) these variations are much slower than the time of a correction loop. Moreover, one can change the value of $\phi_{op}$ directly during the correction to compensate for slight changes. However, in an operational instrument, this problem will be taken into account by design to comply with the stability requirements~\citep{Macintosh08}.

\section{Correction in the focal plane using the self-coherent camera: laboratory performance}
\label{sec:SCCCorrection}

\subsection{Laboratory test bench}
\label{sec:Bench_THD}

We tested the SCC on a laboratory bench at the Observatoire de Paris. A thorough description of this optical bench is given in~\cite{mmas_spie_2010}. We briefly present the main components used in the experiments of the current paper: 

\begin{enumerate}
\item A quasi-monochromatic laser diode emitting at 635nm.
\item A tip-tilt mirror built at LESIA, used to center the beam on the coronagraphic mask \citep{Mas12}. The tip-tilt mirror can also be used in the closed-loop as an off-load for the DM.
\item A Boston Micromachines DM of 32x32 actuators on a square array. Each actuator has a size of 300$\mu m$. We currently use an entrance pupil of 8.1mm and thus 27 actuators across the pupil.
\item An FQPM optimized for 635nm.
\item A Lyot stop with a diameter of 8mm for an entrance pupil of 8.1mm ($98.7\%$ filtering) and in the same plane, reference pupils of variable diameters: 0.3mm ($\gamma = 26.6$), 0.35mm ($\gamma = 22.8$), 0.4mm ($\gamma = 20$) and 0.5mm ($\gamma = 16$) and 0.8mm ($\gamma = 10$).
\item A CCD camera of 400x400 pixels with a readout noise of 16 electrons/pixel and a full well capacity of 13,000 electrons/pixel.
\end{enumerate}

We used the Labview software to control the bench and the DM and applied the closed-loop correction at 20 Hz.

\subsection{Dark holes}
\label{sec:DH}

\begin{figure}
   \begin{center}
   \includegraphics[width=9cm]{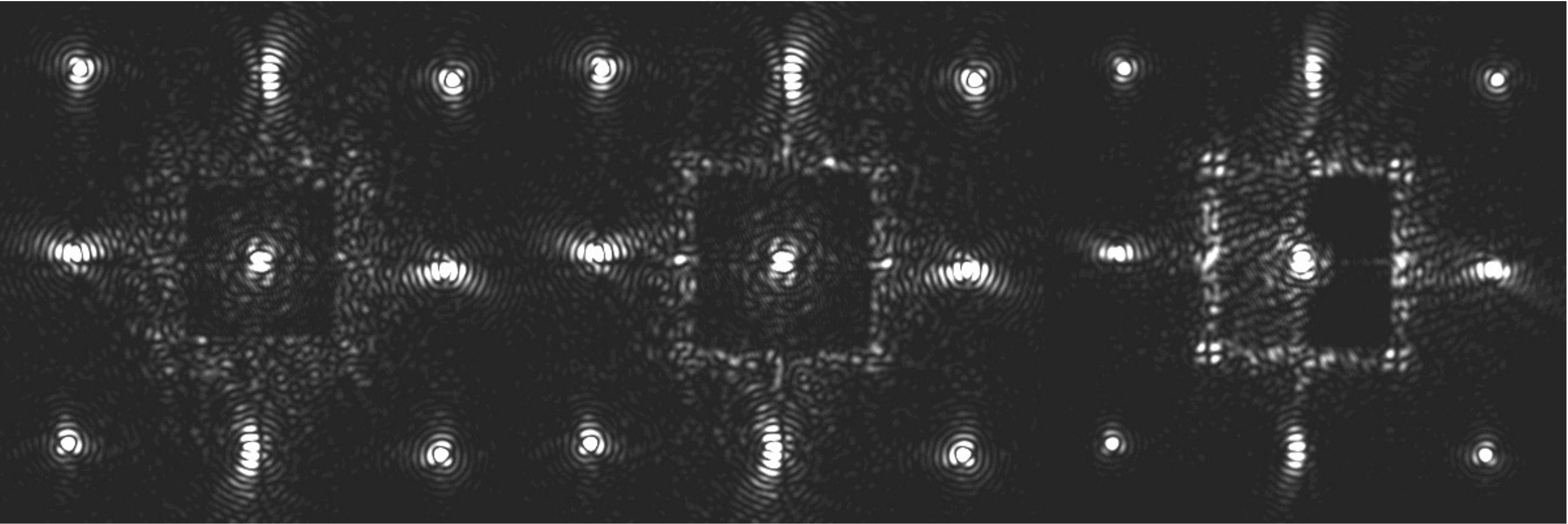} 
   \end{center}
   \caption[three_DH] 
   { \label{fig:three_DH} 
\emph{Dark holes recorded on the laboratory bench for correction with two different sizes of square mask $S_q$: $K_{S_q} = 20.8 \lambda/D_{L}$ (left) and $K_{S_q} = 24.5 \lambda/D_{L}$ (center). The dark hole recorded on the laboratory bench for a correction in phase and amplitude with a square mask of size $K_{S_q} = 24.5 \lambda/D_{L}$ (right). These images use a different intensity scale but the same space scale}}
 \end{figure}  

Owing to the limited number of actuators on the DM, only spatial frequencies lower than the DM cut-off can be corrected for. For a given diameter $D_L$ of the Lyot pupil, the highest frequency attainable for a NxN actuators DM (N actuators across the pupil diameter) is $N\lambda/(2D_L)$ in one of the principal directions of the mirror and $\sqrt{2}N\lambda/(2D_L)$ in the diagonal. The largest correction zone, called dark hole ($\mathcal{DH}$) in~\cite{Malbet95} is the zone $\mathcal{DH}_{max} =[-N\lambda/(2D_P),N\lambda/(2D_P)]\times[-N\lambda/(2D_P),N\lambda/(2D_P)]$ in the image plane. During the numerical process of the SCC image (Figure~\ref{fig:step_estim_fo}), we can decide to reduce the correction to a smaller zone than the one allowed by the number of actuators of the DM. This can be implemented in the SCC correction by multiplying $I-$ by a square mask $S_q$. Modifying Equation~\ref{eq:corr_operator_ac_SCC_opd}, the estimation becomes
\begin{equation}
	\label{eq:psiS_in_p_plan_modif_size}
	\Phi_{est} = gi\mathcal{F}^{-1}\left[\frac{S_q.I_-\exp(-i\phi_{op}^{mes})}{M}\right]\,.P,
\end{equation}
where $S_q$ equals 1 on a square area of $K_{S_q}\lambda/D_{L}$x$K_{S_q}\lambda/D_{L}$ in the center of the image and 0 everywhere else.

Using an SCC with a reference pupil of 0.5 mm ($\gamma = 16$), we applied Equation~\ref{eq:psiS_in_p_plan_modif_size} to estimate the upstream wavefront. We used a square zone to restrain the correction zone to $24.5\lambda/D_{L}$ to optimize the correction of the DM. We built a synthetic interaction matrix as described in Section~\ref{sec:SyntMatr}. The pseudo inverse of $D^{synth}$ was used to control the DM in a closed loop using Equation~\ref{eq:corDM}. The correction loop was closed at 20 Hz for the laboratory conditions and ran for a number of iterations large enough ($j>10$) for the DM to converge to a stable shape. We recorded focal plane images during the control loop. The typical result obtained on the optical bench for this reference and square zone sizes and for phase-only correction is shown in Figure~\ref{fig:three_DH} (center). We also show an image of a $\mathcal{DH}$ obtained with a correction with a square zone of size $K_{S_q} = 20.8 \lambda/D_{L}$ (Figure~\ref{fig:three_DH}, left). A specific study of the size of the correction zone is made in Section~\ref{sec:DH_size}. In Figure~\ref{fig:three_DH}, dark zones represent low intensities. 
The eight bright peaks at the edges are caused by high spatial frequencies due to the print-through of the actuators on the DM surface. These peaks are uncorrectable by nature, but probably do not strongly alter the correction because they are situated at more than $20\lambda/D_{L}$ from the center.

As explained in Section~\ref{sec:phase_minimize}, the correction of phase and amplitude with only one DM is possible by replacing $A_S$ by $A_S^{hermi}$ in Equation~\ref{eq:estim_As_and_gain}. With Equation~\ref{eq:hermi_def}, we similarly define the  hermitian function $I_-^{hermi}$ from $I_-$. Using Equation~\ref{eq:AS_function_I} and the assumption that $|A_R^*|^2$ is an Airy pattern, a phase and amplitude correction is therefore possible by replacing $I_-$ by $I_-^{hermi}$ in Equation~\ref{eq:psiS_in_p_plan_modif_size}. This correction allows one to go deeper in contrast but limits the largest possible dark hole to half of the focal plane: $\mathcal{DH}_{max}^+ =[0,N\lambda/(2D_L)]\times[-N\lambda/(2D_L),N\lambda/(2D_L)]$. On this half plane, we can also choose to reduce the correction to a smaller zone. A resulting dark hole is presented in Figure~\ref{fig:three_DH} (right) for $K_{S_q} = 20.8 \lambda/D_{L}$.

\subsection{SCC performance}
\label{sec:optim}

In this section, we present contrast results obtained on the laboratory bench for phase-only correction and for amplitude and phase correction. We used a reference pupil of 0.5 mm ($\gamma = 16$) to estimate the upstream wavefront and a square zone of size $K_{S_q} = 24.5\lambda/D_{L}$ to optimize the correction of the DM.

\subsubsection{Phase-only correction}
\label{sec:optimPhase}

\begin{figure}
   \begin{center}
   \includegraphics[width=8cm]{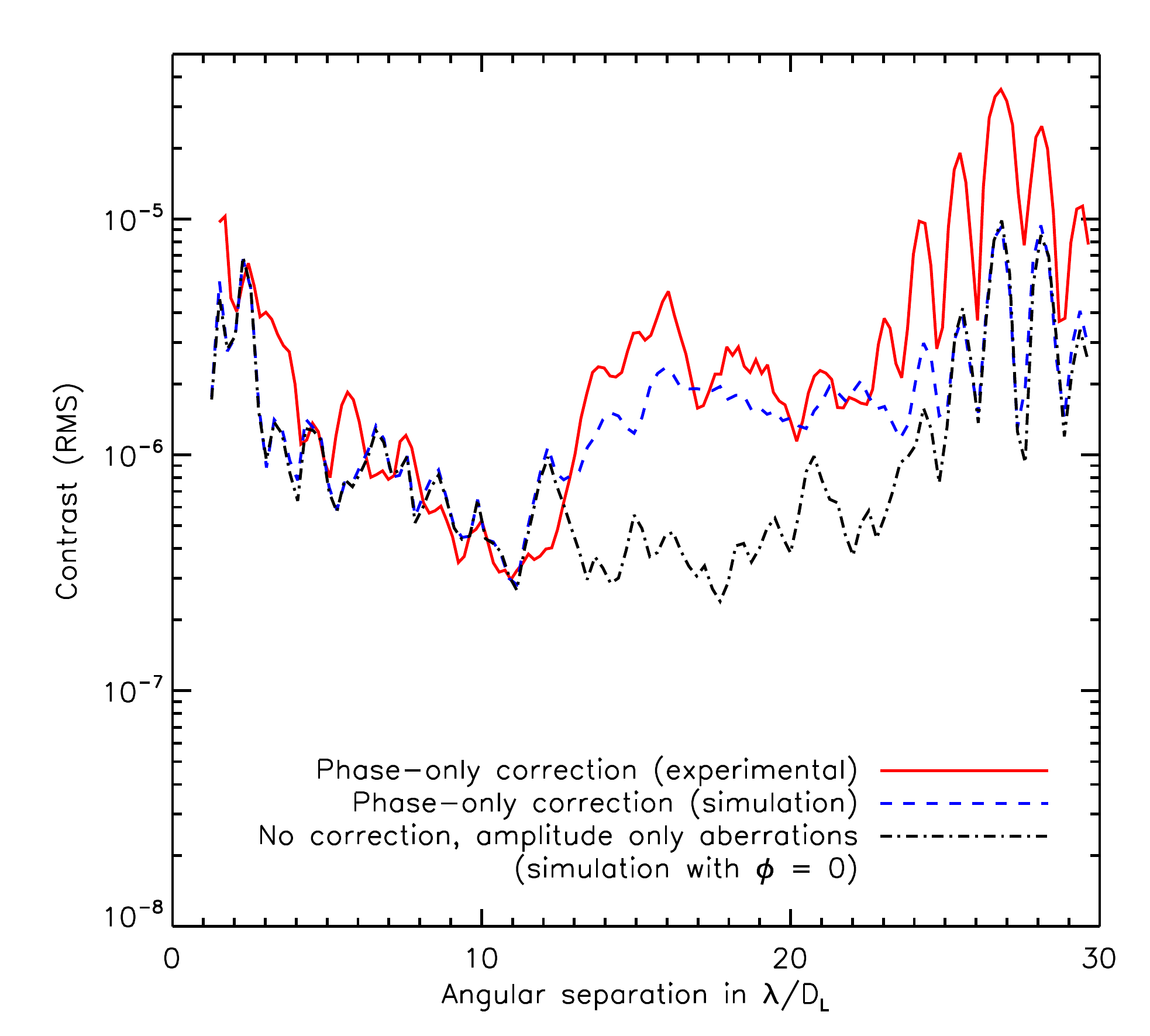} 
   \end{center}
   \caption[best_result] 
   { \label{fig:best_result_ph} 
\emph{Radial profiles of the azimuthal standard deviation (in RMS) of the intensities in the focal plane typically obtained with this method for phase-only correction, for simulation (blue dashed line) and laboratory bench result (red solid line), for $\gamma = 16$ and a square zone of size $K_{S_q} = 24.5\lambda/D_{L}$. We also plot in this graph the simulation of the focal plane obtained using the amplitude aberrations recorded and no phase aberrations (black dash-dotted line).}}
   \end{figure}  

The speckles near the FQPM transitions are brighter than those in other parts of the $\mathcal{DH}$. Moreover, the contrast in these region is not relevant, because the image of a planet located on a transition would be distorted and strongly attenuated. Therefore, for phase-only correction, we chose to measure the radial profile of the SCC image only on the points (x,y) which verify
\begin{equation}
\label{eq:far_from_trans}
   \left \{
   \begin{array}{c}
      x \in [-20\lambda/D_{L};-1\lambda/D_{L}] \cup [1\lambda/D_{L};20\lambda/D_{L}] \\
      y \in [-20\lambda/D_{L};-1\lambda/D_{L}] \cup [1\lambda/D_{L};20\lambda/D_{L}].
   \end{array}
   \right .
\end{equation}
We calculated the profiles by normalizing the intensities by the highest value of the PSF measured through the Lyot pupil and without coronagraphic mask. In practice, we moved the source away from the coronagraph transitions to measure this PSF. In the following figures, the distances to the center are measured in $\lambda/D_L$. Figure~\ref{fig:best_result_ph} shows the radial profile of the azimuthal standard deviation of the intensities obtained in phase-only correction in the focal plane zone described in Equation~\ref{eq:far_from_trans}. The detection level reaches a contrast level of $10^{-6}$ between 6 and 12 $\lambda/D_L$ and $3.10^{-7}$ at 11 $\lambda/D_L$. As shown in Figure~\ref{fig:three_DH} (center), speckles are still present in the dark area. Since we only corrected for the phase, we can suspect amplitude effects. 

\begin{figure}
   \begin{center}
   \begin{tabular}{c}
   \includegraphics[width=4cm]{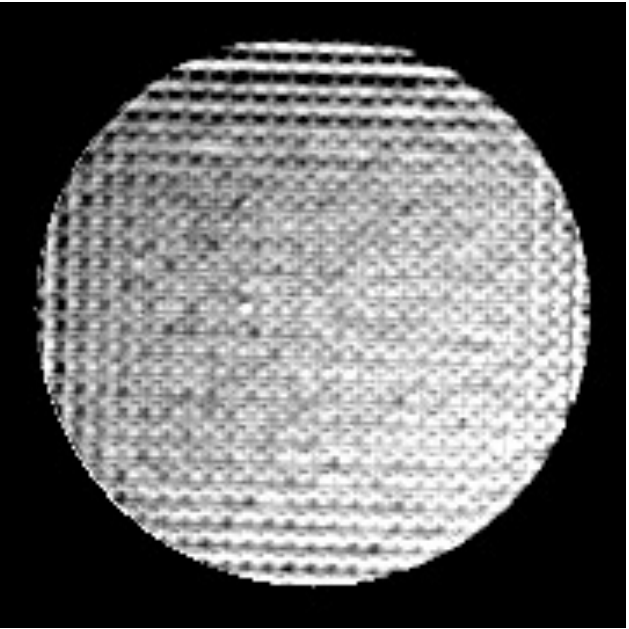}  
   \end{tabular}
   \end{center}
   \caption[image_pup] 
   { \label{fig:image_pup} 
\emph{Pupil illumination recorded on the laboratory bench.}}
   \end{figure}

To estimate the amplitude aberration level, we recorded the pupil illumination on the optical bench without coronagraph, shown in Figure~\ref{fig:image_pup}. The amplitude defect level is estimated to be about 10\% RMS in intensity. The period of the actuator pitch clearly appears in this pupil image. Due to vignetting effects by the focal coronagraphic mask, these high-frequency structures of the DM surface create illumination variations across the pupil. The first effect of these high-frequency aberrations are bright speckles outside the corrected zone (mostly on the eight bright peaks). The second effect is more critical for our purpose. Because the level of high-frequency amplitude errors varies across the pupil, it creates low-frequency amplitude aberrations, which induce bright speckles in the center of the correction zone. 

To compare the level of the recorded speckles with the one expected using amplitude and phase errors, we simulated the expected focal plane image. We used the amplitude aberrations deduced from the intensity measurement on the laboratory bench (Figure~\ref{fig:image_pup}). From these amplitude aberrations, we first simulated the the focal plane without phase errors (just amplitude errors). The profile of this focal plane is plotted in Figure~\ref{fig:best_result_ph} with a black dot-dashed line. We then simulated a phase-only correction, assuming initial phase aberrations of 16 nm RMS over the pupil, and a power spectral density (PSD) in $f^{-2}$ where $f$ is the spatial frequency. These simulation results (blue dashed line) are compared to the experimental measurement (red line) in Figure~\ref{fig:best_result_ph}. The level and shape of the two curves are very similar. They show the same structure around 27$\lambda/D_L$, due to the eight bright peaks created by amplitude aberrations. These curves inside the $\mathcal{DH}$ match the simulation of the focal plane without amplitude aberrations. It seems that in phase-only correction, we corrected all phase aberrations and that we are only limited by amplitude errors.

\subsubsection{Phase and amplitude correction}
\label{sec:optimAmpl}

The simulation without amplitude errors (only phase aberrations) shows that a contrast level of $10^{-10}$ can be reached, as previously shown in~\cite{Galicher10}. Since the amplitude errors set the limits of our phase-only corrections, we aim to correct both phase and amplitude at the same time. However, with only one DM, the corrected zone is smaller by half, as shown in Figure~\ref{fig:three_DH} (right). Therefore, the radial profile measurement zone becomes
\begin{equation}
\label{eq:far_from_trans_corr_ampl}
   \left \{
   \begin{array}{c}
  	 x \in  [1\lambda/D_{L};20\lambda/D_{L}] \\
     y \in [-20\lambda/D_{L};-1\lambda/D_{L}] \cup [1\lambda/D_{L};20\lambda/D_{L}].
   \end{array}
   \right .
\end{equation}

\begin{figure}
   \begin{center}
   \includegraphics[width=8cm]{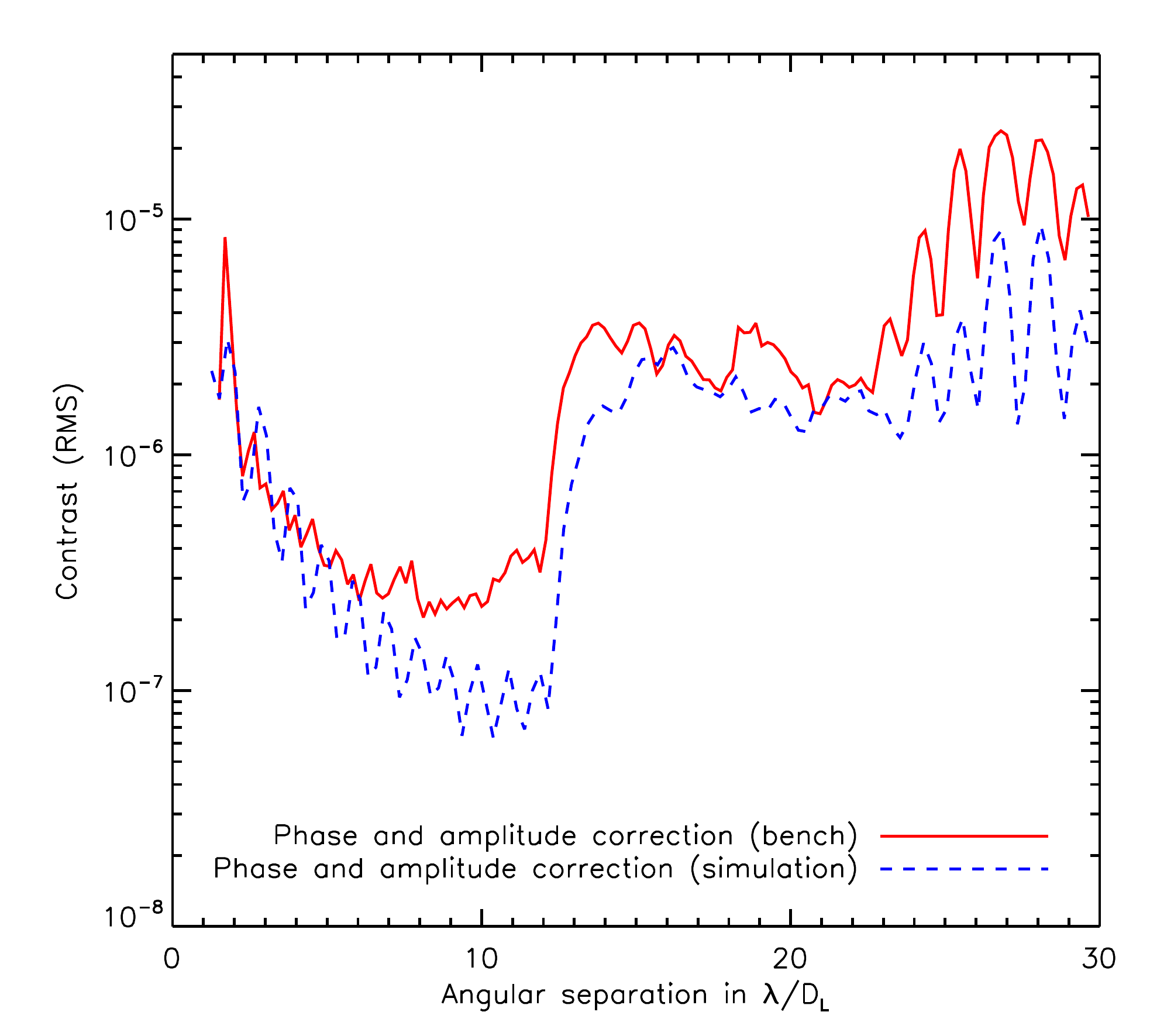} 
   \end{center}
   \caption[best_result] 
   { \label{fig:best_result_amp} 
\emph{Radial profiles of the azimuthal standard deviation (in RMS) of the intensities in the focal plane typically obtained with this method for phase and amplitude correction, for simulation (blue dashed line) and laboratory bench result (red solid line), for $\gamma = 16$ and a square zone of size $K_{S_q} = 24.5\lambda/D_{L}$.}}
   \end{figure}

The results for this correction are plotted in Figure~\ref{fig:best_result_amp} as a dashed blue line for the simulations and as a red line for the laboratory bench results. When correcting for the phase and amplitude aberrations, we obtain contrasts better than $10^{-6}$ between 2$\lambda/D_L$ and 12$\lambda/D_L$, and better than $3.10^{-7}$ between 7$\lambda/D_L$ and 11$\lambda/D_L$. This is an improvement compared to the phase-only correction. The simulated profiles match the laboratory results from 0 to 8 $\lambda/D_L$ and outside of the $\mathcal{DH}$. 

Between 8 and 12 $\lambda/D_L$, the experimental correction shows a plateau at $3.10^{-7}$, while the simulation correction goes deeper. This plateau is a distinctive feature of a limitation caused by the low dynamic range of the detector (our CCD camera has a full well capacity of 13,000 electrons/pixels for a readout noise of 16 electrons/pixels). This is confirmed by the last images of the loop which show speckle levels below the readout noise between 8 and 12 $\lambda/D_L$: the speckles beyond the readout noise are not visible and thus beyond correction. However, this problem can be solved by using a detector with a better dynamic range.

The number of incoming photons from the observed source is a critical problem of any speckle-correction technique: the speckles can only be corrected for to a certain level of contrast if the source is bright enough for them to be detected above photon and detector noise at these levels. Although we can correct in a closed loop at 20 Hz in the laboratory, the correction rate in a real telescope observation will be limited by the shortest exposure time necessary. This shortest exposure time depends on several parameters such as stellar magnitude, observational wavelengths, telescope diameter, or dynamic range of the camera.

The contrast level in the numerical simulation is limited to $10^{-7}$. This is due to the high-amplitude defects ($10\%$ in intensity) introduced by the DM in the pupil. Indeed, the bright speckles of the uncorrected half-area diffract their light into the corrected half-area. This limit, independent of the estimation method \citep{Giveon06,Galicher10}, may be lowered by the introduction of a second DM on the optical bench \citep{Pueyo10}.

In the next sections (Section~\ref{sec:DH_size}~and~\ref{sec:SizeRef}), we study the influence of different parameters on the SCC performance.
	
\subsection{Size of the corrected zone}
\label{sec:DH_size}

In this section, we compare the performance for different sizes of the square zone $S_q$. Using the modified estimator introduced in Equation~\ref{eq:psiS_in_p_plan_modif_size}, and for different square zone sizes $K_{S_q}$, we experimentally closed the loop and recorded images after convergence. In these tests, we used $N =27$ actuators across the pupil diameter and $\gamma=16$, with phase-only correction. As explained in Section~\ref{sec:DH}, for this number of actuators, we have $\mathcal{DH}_{max} = [-26.6\lambda/(2D_L),26.6\lambda/(2D_L)]\times[-26.6\lambda/(2D_L),26.6\lambda/(2D_L)]$ (as $D_L/D_P = 8/8.1$). We tested the case ($K_{S_q} = \infty$) and three others: $K_{S_q} = 26.4 \lambda/D_{L}$, which is only slightly smaller than size of the largest $\mathcal{DH}$ and two smaller square zones ($K_{S_q} = 20.8 \lambda/D_{L}$ and $K_{S_q} = 24.5 \lambda/D_{L}$). The images obtained in the last two cases can be seen in Figure~\ref{fig:three_DH}:  $K_{S_q} = 20.8 \lambda/D_{L}$ (left) and $K_{S_q} = 24.5 \lambda/D_{L}$ (center).

\begin{figure}
   \begin{center}
   \begin{tabular}{c}
   \includegraphics[width=8cm]{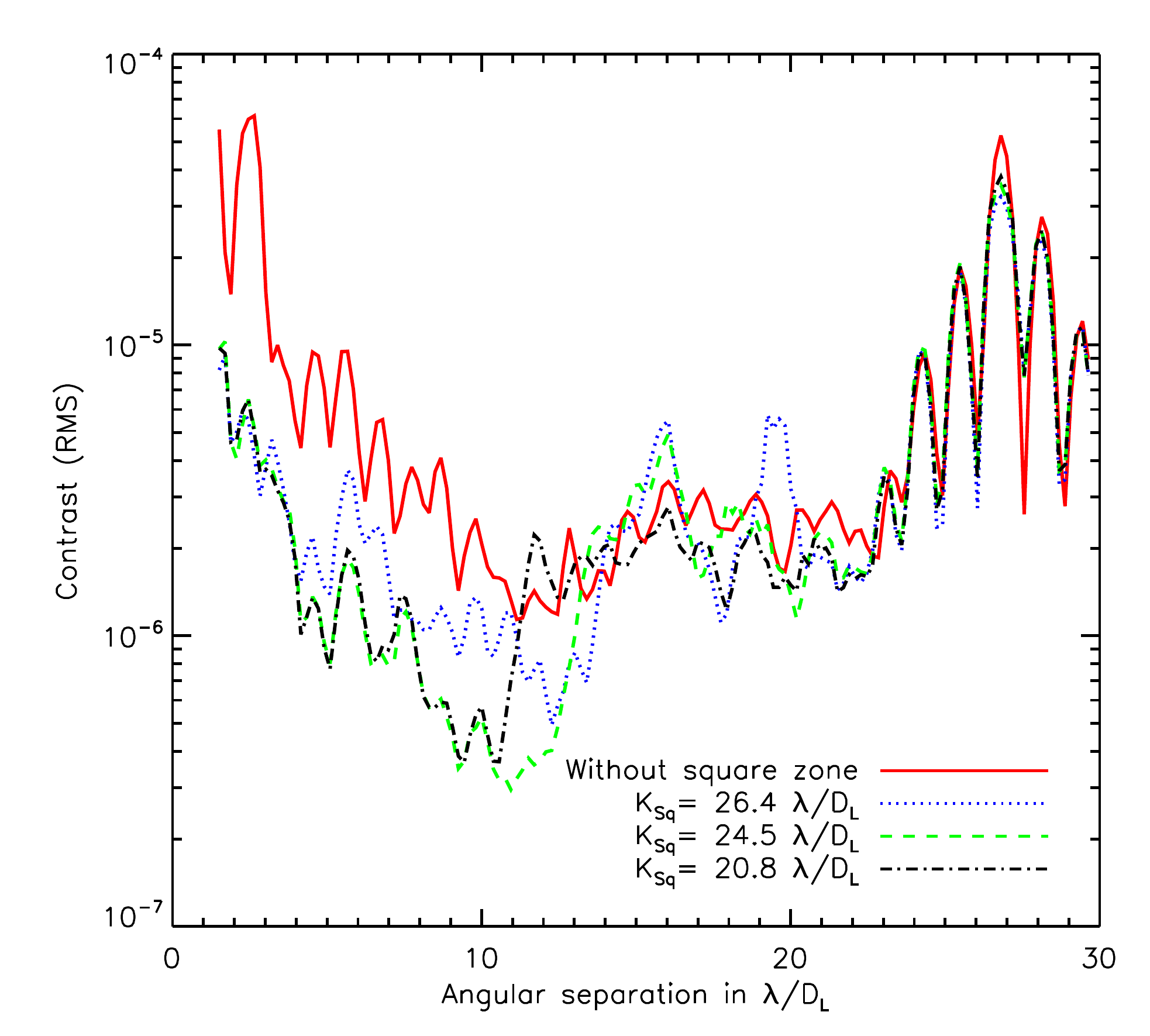}  
   \end{tabular}
   \end{center}
   \caption[compar_sq_zone] 
   { \label{fig:compar_sq_zone} 
\emph{Experimental radial profile comparison of dark holes obtained on the test bench without square zone (directly using the estimator described in Equation~\ref{eq:corr_operator_ac_SCC_opd}) (red, solid) and with square zones of different side lengths: $K_{S_q} = 26.4 \lambda/D_{L}$ (blue, dotted), $K_{S_q} = 24.5 \lambda/D_{L}$ (green, dashed) and $K_{S_q} = 20.8 \lambda/D_{L}$ (black, dot-dashed). These phase-only corrections were achieved with a $\gamma=16$ reference pupil. The intensities are normalized by the highest value of the PSF obtained without coronagraphic mask.}}
   \end{figure}

Figure~\ref{fig:compar_sq_zone} presents the radial profiles of the focal planes obtained on the laboratory bench, normalized by the highest value of the PSF obtained without coronagraphic mask. 

The red, solid curve shows the result for $K_{S_q} = \infty$, without square zone. The blue dotted line represents the result of a square mask of size $K_{S_q} = 26.4 \lambda/D_{L}$, which is only slightly smaller than the actual cut-off frequency of the DM. In this case, we prevented the correction of speckles outside of the $\mathcal{DH}$ and obtained a great improvement inside the $\mathcal{DH}$ (0 to 13.5 $\lambda/D_{L}$) and a small depreciation outside (13.5 to 15.5 $\lambda/D_{L}$). Using a smaller correction zone ($K_{S_q} = 24.5 \lambda/D_{L}$ green dashed line) still improves the correction but to the detriment of the size of the $\mathcal{DH}$ (the contrast starts to rise around $12\lambda/D_{L}$). Finally, we see that a smaller square zone ($K_{S_q} = 20.8 \lambda/D_{L}$, black, dot-dashed) produces a smaller but not shallower $\mathcal{DH}$. 

Going from $K_{S_q} = \infty$ to $K_{S_q}  = 24.5 \lambda/D_{L}$, the contrast in the $\mathcal{DH}$ progressively deepens. This is because correcting fewer of the highest frequencies with a constant number of actuators, we free degrees of freedom. However, for $K_{S_q} < 25.5 \lambda/D_{L}$, the contrast level does not improve because we reach the level of the speckles created by the amplitude aberrations. Additional shrinking would only reduce the size of the $\mathcal{DH}$. Thus, the reduction of the corrected zone in the wavefront estimation greatly improves the correction performance (up to a factor 10) with only a small reduction of the $\mathcal{DH}$ size. This effect was described in~\cite{BordeTraub06} using 1D simulations. 

It is important to note that this improvement does not come from the phenomenon of aliasing in the estimation \citep{Poyneer04}. Indeed, only the correction is enhanced by this process, because the estimation remained unchanged. The wavefront estimation with the SCC is only limited in frequency by the size of the reference PSF: we can estimate speckles as long as the reference flux is not null, \emph{i.e.}, as long as the speckles are fringed. In most cases (see next section), the first dark ring of the reference PSF is larger than the correction zone and the frequencies inside the PSF's first dark ring are well estimated. 

\subsection{Size of the reference pupil}
\label{sec:SizeRef}

In this section, we study the effect of the size of the reference pupil on the performance of the SCC. In the previous sections, we used two assumptions on the size of the reference pupil. First, in Section~\ref{sec:data_exctract}, we assumed a reference pupil small enough to consider that the influence of the aberrations inside such a reference pupil is negligible. Simulations showed that even for small $\gamma$, the level of aberrations in the reference pupil is very low and uncorrelated to the level of aberrations in the entrance pupil. Second, in Section~\ref{sec:SCC_estim}, we assumed a reference pupil small enough to consider $A^*_R$ constant over the correction zone  in the focal plane. As previously mentioned, the highest frequency attainable by the DM is $\sqrt{2}N\lambda/(2D_{L})$. Using the first assumption, $|A^*_R|^2$ is a perfect PSF whose first dark ring is located at $1.22 \lambda\gamma/D_{L}$. Thus, $A_R^*$ is roughly constant over the $\mathcal{DH}$ if
\begin{figure}
   \begin{center}
   \begin{tabular}{c}
   \includegraphics[width=8cm]{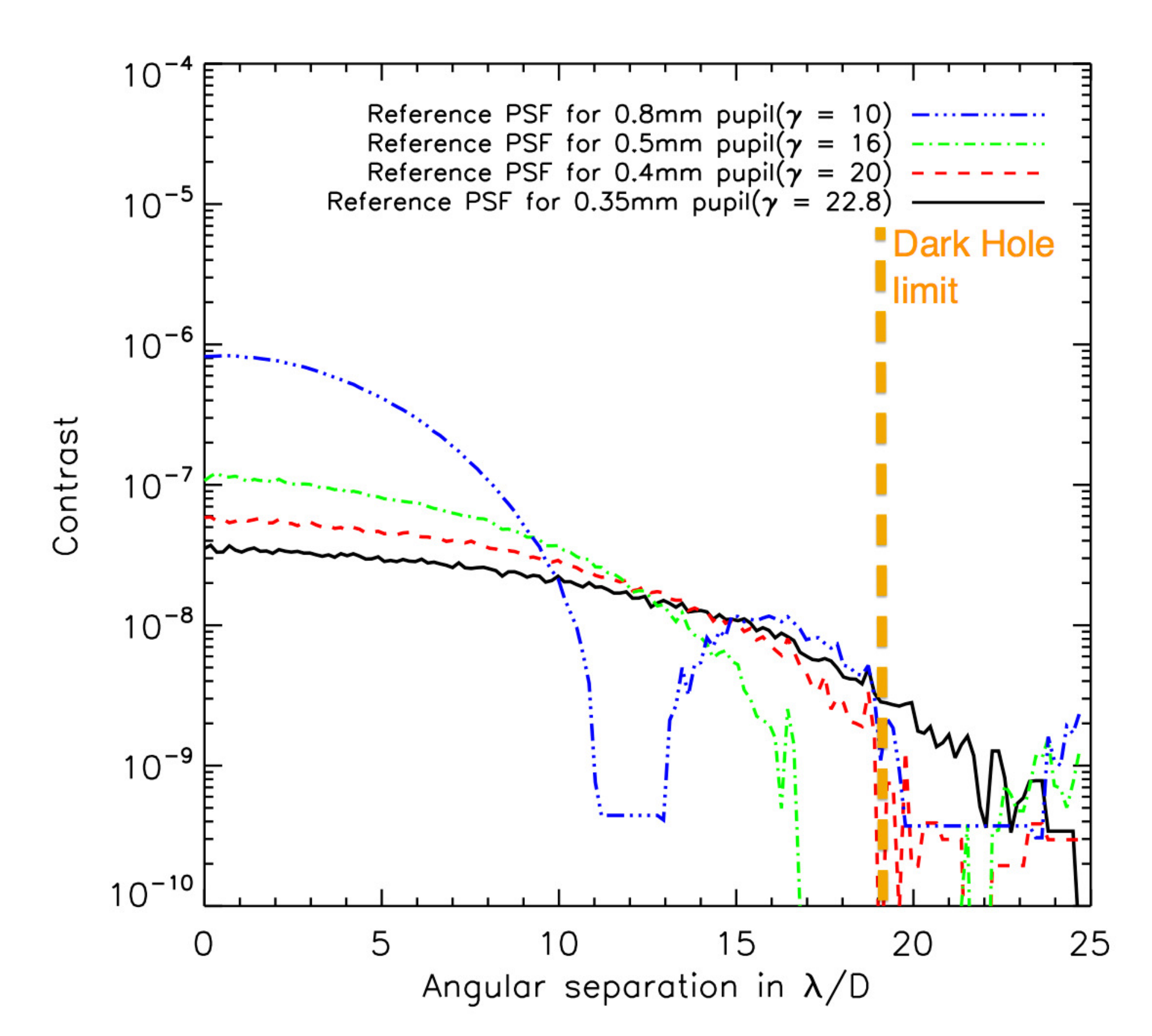}
   \end{tabular}
   \end{center}
   \caption[psf_refs] 
   { \label{fig:psf_refs} 
\emph{Experimental radial profiles of the PSFs for reference pupils from $\gamma=10$ to $\gamma=22.8$ recorded on the optical bench. The distance to the center is in $\lambda/D_{L}$. These reference PSFs are normalized by the highest value of the Lyot PSF obtained without coronagraphic mask. The vertical line correspond to the frequency cut-off for $N = 27$ actuators in the entrance pupil ($\sqrt{2}N\lambda/(2D_{L})$)}}
   \end{figure}  
\begin{equation}
	\label{eq:airy_in_dh}
\ \ 1.22\gamma > N/\sqrt{2}.
	\end{equation}
For $N = 27$ actuators in the entrance pupil, Equation~\ref{eq:airy_in_dh} reads $\gamma > 15.6$. In Figure~\ref{fig:psf_refs}, we plot the radial profiles of $|A_R|^2$ recorded on the optical bench for $\gamma$ from $10$ to $22.8$. We observe a wide range of intensity levels for different reference pupils (from $10^{-6}$ for $\gamma = 10$ to $3.10^{-8}$ for  $\gamma = 22.8$). A reference pupil with $\gamma=10$ (blue, solid) does not satisfy Equation~\ref{eq:airy_in_dh}, and the first ring of its PSF is inside the correction zone (vertical orange dashed line). We test this case independently in Section~\ref{sec:large_ref}. The other reference pupils are studied in Section~\ref{sec:small_ref}. 

\subsubsection{Impact of small reference pupils}
\label{sec:small_ref}

The size of the reference pupil can influence the correction in two different ways: it changes the signal-to-noise ratio (S/N) on the fringes and modifies the flatness of the reference PSF over the correction zone. We develop these effects in this order in this section. 

The S/N on the fringes is critical, because $I_-$ can only be retrieved with well-contrasted fringes. The S/N is directly related to the reference pupil size. Using Equation~\ref{eq:focal_plane}, we deduce that the peak-to-peak amplitude of the fringes in the focal plane is $2|A_S||A_R|$. Thus, if $|A_S|$ and $|A_R|$ are expressed in photons, and assuming only photon and read-out noise, the S/N can be written as
\begin{equation}
	\label{eq:S/N}
\ \ S/N \simeq \dfrac{2|A_S||A_R|}{\sqrt{|A_S|^2+|A_R|^2+ \sigma_{cam}^2}},
	\end{equation}
where $\sigma_{cam}$ is the standard deviation of the detector noise in photons. A higher S/N allows a better estimate of the speckle complex amplitude and thus, a better correction of the aberrations. One can notice that this S/N can be simplified depending on the relative values of its different terms. We quickly study the following cases:
\begin{itemize}
\item if $|A_S| \approx |A_R| \ll \sigma_{cam}$, $S/N \rightarrow 0$. In this case, the correction is limited by the dynamic range.
\item if $\sigma_{cam} \ll  |A_S|$ and $|A_R| \ll |A_S|$,  $S/N \sim 2|A_R|$. Initial case, at the beginning of the correction, when the Lyot pupil is a lot brighter than the reference pupil. The S/N is only a function of $|A_R|$.
\item if $\sigma_{cam} \ll  |A_R|$ and $|A_S| \ll |A_R|$,  $S/N \sim 2|A_S|$. The S/N is decreasing with deepening correction. The reference brightness is not important.
\end{itemize}
Equation~\ref{eq:S/N} shows that this S/N is an increasing function of $|A_R|$, but for deep corrections ($|A_S| \ll |A_R|$), the impact of the size of the reference is probably very weak.

The second effect is due to the assumption of a constant reference PSF over the correction zone. Variations of $A_R^*$ in the correction zone distort the wavefront estimation. This effect advocates for small reference pupils (large $\gamma$): a reference pupil of $\gamma = 16$ generates an $A_R^*$ that varies from $1$ to $0.03$ inside a correction zone of $27$x$27\lambda/D_{L}$. For this reference pupil, the fringe intensity is weaker at the edges of the $\mathcal{DH}$. Therefore, the estimate is less accurate at these locations.

Using simulation tools, where we can change the camera and photon noise easily, we were able to isolate these two different effects and analyzed their influence on the performance of the instrument separately. A more detailed study has previously been presented in~\cite{MazoyerSPIE12}. 

\begin{figure}
   \begin{center}
    \begin{tabular}{c}
   \includegraphics[width=8cm]{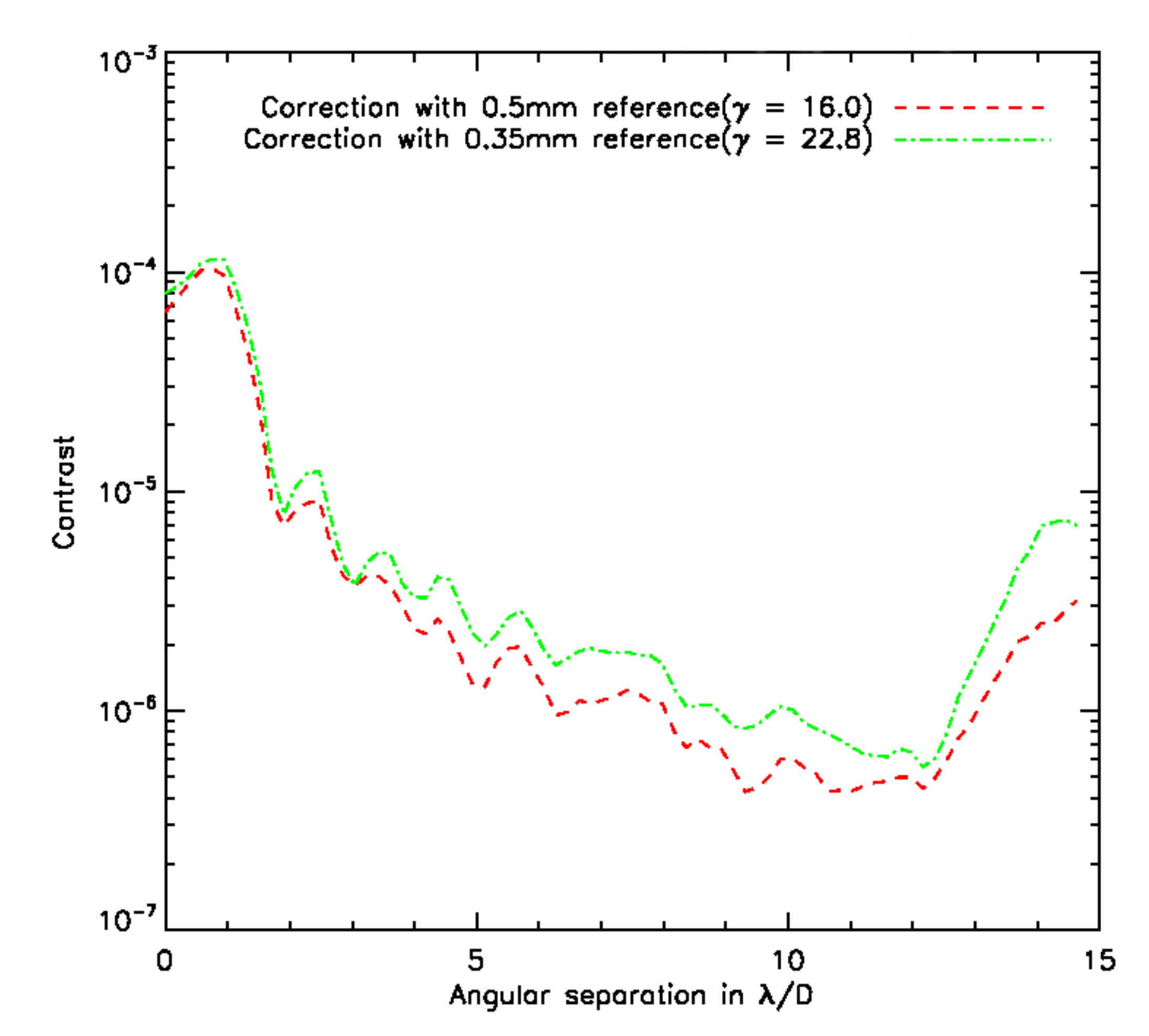}
   \end{tabular}
   \end{center}			
   \caption[test_bench_size] 
   { \label{fig:test_bench_size} 
\emph{Radial profiles obtained on the laboratory bench for two different reference pupils ($\gamma = 16$ and $\gamma = 22.8$) with $N = 27$ actuators and $K_{S_q} = 24.5 \lambda/D_{L}$ for the size of the corrected zone. These contrasts are normalized by the highest value of the PSF obtained without coronagraphic mask.}} 
      \end{figure}

Here, we experimentally tested the influence of the reference size. We used 27 actuators across the pupil diameter and $K_{S_q} = 24.5 \lambda/D_{L}$ with phase-only correction. Figure~\ref{fig:test_bench_size} shows the radial profiles of the SCC image in RMS obtained on the laboratory bench for different reference pupils ($\gamma = 16$ and $\gamma = 22.8$), normalized by the highest value of the PSF obtained without coronagraphic mask. These results show that a large reference pupil ($\gamma = 16$) is preferable, even at the edge of the $\mathcal{DH}$, where the reference PSF for $\gamma = 16$ is fainter than the reference PSF for $\gamma = 22.8$. Comparing the contrast levels obtained in this figure with those in Figure~\ref{fig:psf_refs} for all reference pupils, we deduce that we are still in the case $|A_R| \ll |A_S|$. Deeper corrections would normally depend less on the size of the reference pupil.

When we use the SCC as a planet finder there is another impact to consider: detection is possible only if the planet intensity is higher than the photon noise of the reference pupil. This effect advocates for small reference pupils. A trade-off study of the reference size is needed depending on expected planet intensity and the actual contrast that can be achieved. A more complete study of the noise in the SCC estimation is given in~\cite{Galicher10}.

\subsubsection{Effect of large reference pupils}
\label{sec:large_ref}

\begin{figure}
   \begin{center}
   \begin{tabular}{c}
   \includegraphics[width=2.7cm]{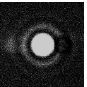}
   \includegraphics[width=2.7cm]{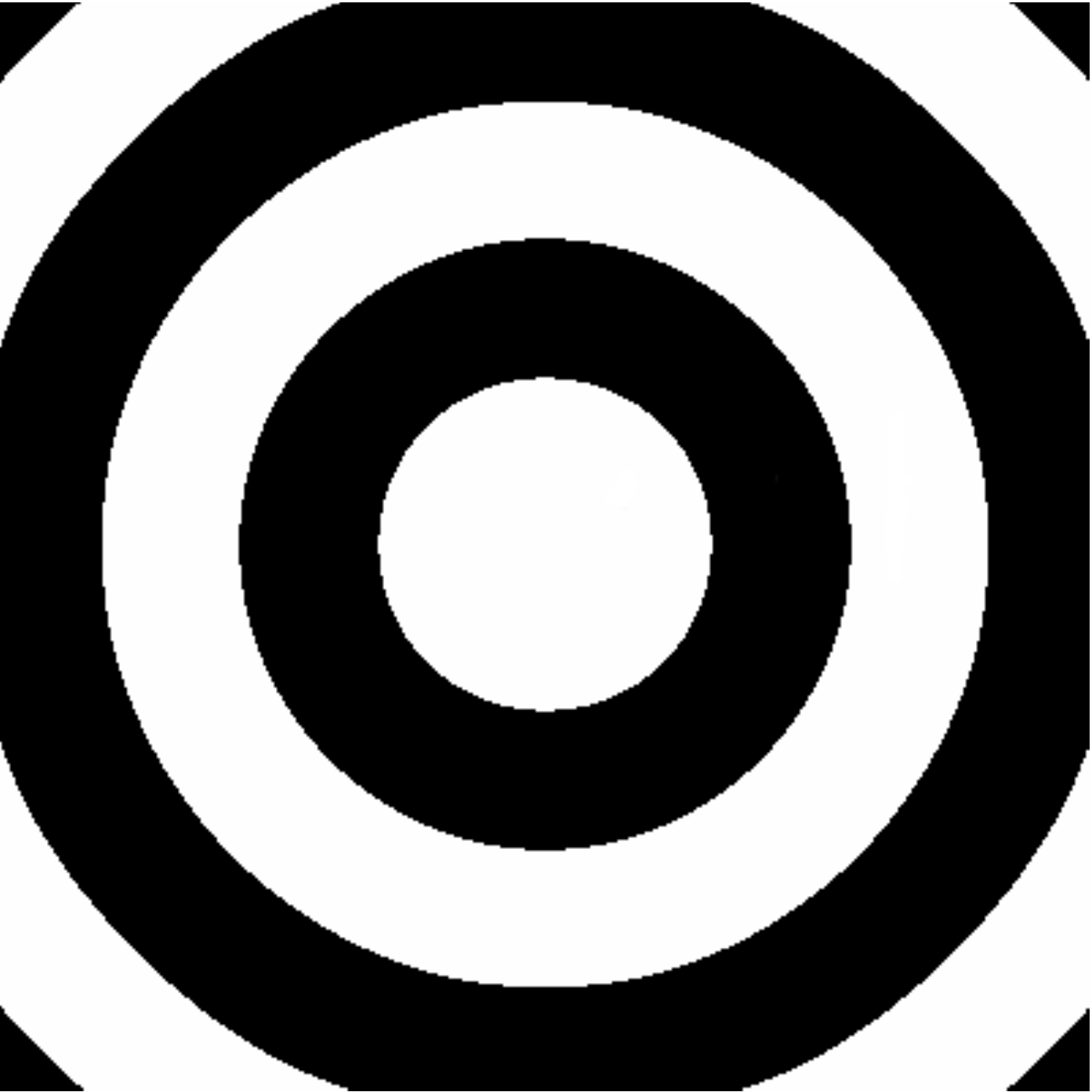}
   \includegraphics[width=2.7cm]{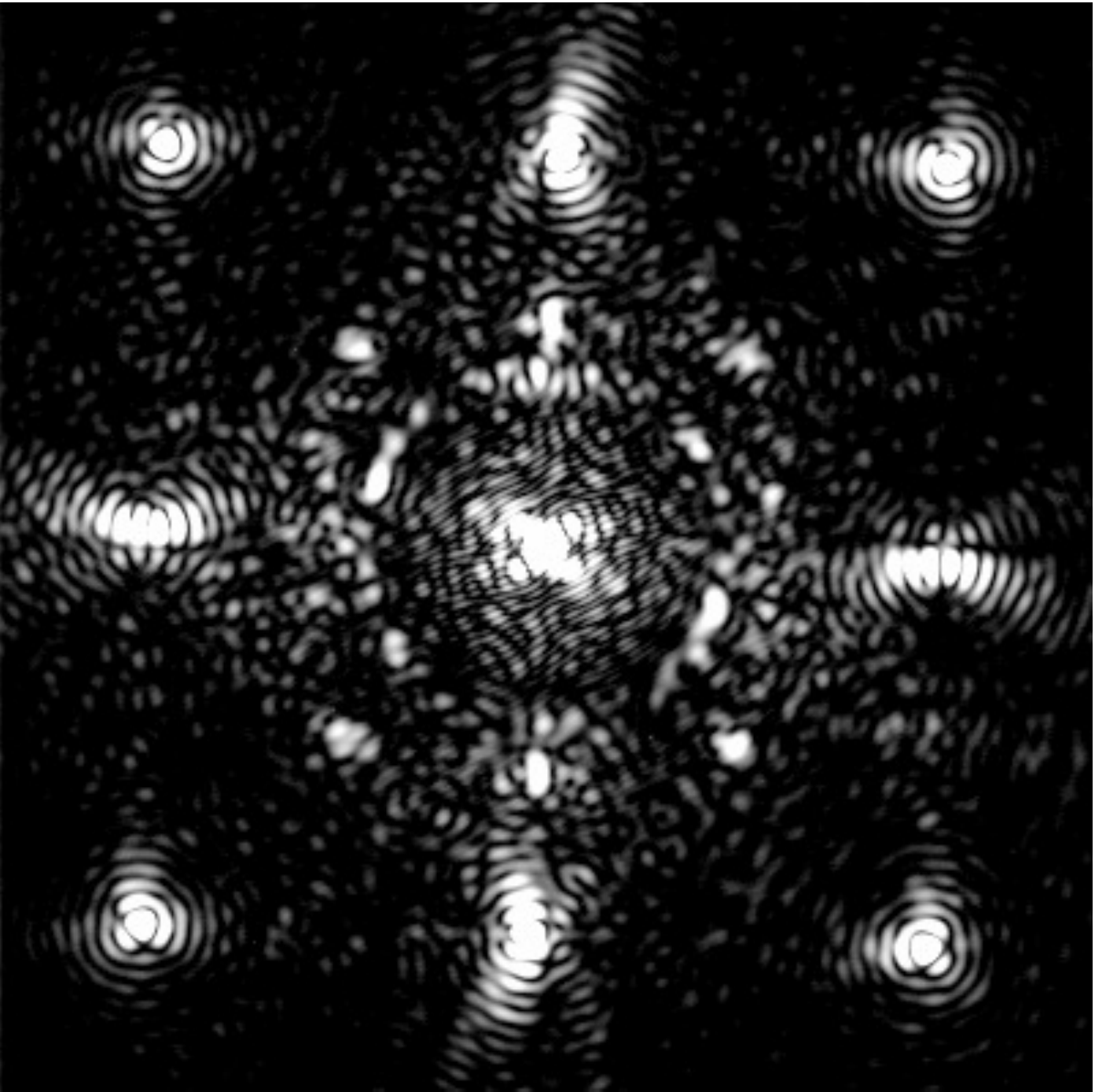}
   \end{tabular}
   \end{center}
   \caption[Correctionref08] 
   { \label{fig:Correctionref08} 
\emph{PSF of the 0.8mm reference pupil ($\gamma = 10$) (right). From this PSF we constructed the sign mask (center). The white zones are uniform and equal to 1 and the black zones are equal to -1. Multiplying $I_{-}$ by this mask, the correction can be achieved (right) for this reference pupil.}}
   \end{figure}

In this section, we experimentally prove that we can still achieve a correction inside the $\mathcal{DH}$ using a reference pupil that does not satisfy Equation~\ref{eq:airy_in_dh} by modifying the phase estimator. This correction has previously been simulated in~\cite{Galicher10}. $A_{R}^*$ is still considered as the complex amplitude of a perfect PSF, but we cannot consider it uniform anymore over the $\mathcal{DH}$. First, the speckles in the first dark ring of this PSF are not fringed, because the reference PSF intensity is null at this location. The wavefront errors that produce these speckles are not estimated and are thus not corrected for. Second, the sign of $\Re[A_{R}^*]$ and $\Im[A_{R}^*]$ changes between the first and the second dark ring (\emph{i.e.}, between $1.22$ and $2.23 \lambda/D_{L}$). These speckles are fringed and we can estimate the wavefront errors that produce them when we consider the sign change. Hence, when Equation~\ref{eq:airy_in_dh} is not satisfied, instead of $A_{R}^*$ constant, we assume $|A_{R}^*|$ constant and change the sign of $A_{R}^*$ over the correction zone. We now estimate
\begin{equation}
	\label{eq:psiS_in_p_plan_modif_sgn}
		\Phi_{est} = i\mathcal{F}^{-1}\left[\frac{Sign[\Re[A_{R}^*]].I_-\exp(-i\phi_{op}^{mes})}{M}\right]\,.P,
		\end{equation}
where $Sign[\Re[A_{R}^*]]$, is the sign of the real part of $A_{R}^*$. This function is represented in Figure~\ref{fig:Correctionref08} (center).

In practice, to achieve the correction with this reference pupil, we multiplied $I_-$ by the mask in Figure~\ref{fig:Correctionref08} (center), where the white zones (the black zones) are constant and equal to $1$ ($-1$). To build this mask, we recorded the reference PSF (Figure~\ref{fig:Correctionref08}, left). From this PSF, we were able to find the dark rings of the complex amplitude. We were able to build the sign of the real part of the complex amplitude. 

The tests on the optical bench were conducted using the 0.8mm reference pupil ($\gamma = 10$) and the process described in Section~\ref{sec:optim}. We used no square zone. The resulting $\mathcal{DH}$ is presented in Figure~\ref{fig:Correctionref08} (right). We distinctly see the first reference ring at $1.22 \lambda/D_{R}$. As expected, the speckles on this ring are not corrected for, because they are not fringed. Nevertheless, apart from this ring, the whole $\mathcal{DH}$ is corrected. Although correction with a large reference pupil is possible, the level of speckle suppression is much lower (better contrast) than with smaller reference pupils (higher $\gamma$), because the speckles of the uncorrected dark ring diffract their light into the corrected zone \citep{Galicher10, Giveon06}.

We showed in Section~\ref{sec:small_ref} that the SCC used with a reference pupil that obeys Equation~\ref{eq:airy_in_dh} shows a better performance. However, some cases (many aberrations due to an unknown initial position of the DM, for example) may require the use of large reference pupils that produce highly contrasted fringes even with very aberrated wavefronts. The correction can then be initiated by correcting for low spatial frequencies (usually dominating the wavefront errors). Finally, the large reference is replaced with a smaller reference (which satisfies Equation~\ref{eq:airy_in_dh}) to correct higher frequencies and reach better contrast levels.

\section{Conclusion}
\label{sec:ccl}

In Section\ref{sec:coro_model}, we used Fourier optics to model the propagation of light through a coronagraph. We then proposed a method for estimating phase and amplitude aberrations in the entrance pupil from the complex electric field measured in the focal plane after a four-quadrant phase mask coronagraph. We used this model to correct phase and amplitude aberrations in a closed loop using a DM in the pupil plane, even for a Lyot pupil smaller than the entrance pupil. 

We implemented this technique, associated with a self-coherent camera as a  focal plane wavefront sensor. We corrected for phase and amplitude aberrations in a closed loop which led to speckle suppression in the central area of the focal plane (called dark hole).

We tested these methods on a laboratory bench where we were able to close the loop and obtain a stable correction at 20 Hz. When correcting for phase aberrations only , we obtained contrast levels (RMS) better than $10^{-6}$ between 6 and 12 $\lambda/D_L$ and $3.10^{-7}$ at 11 $\lambda/D_L$. We proved that we corrected for most phase aberrations in the dark hole and that the contrast is limited by high amplitude aberrations (10\% RMS in intensity) induced by the DM. When correcting for the phase and amplitude aberrations using one DM, we obtained contrast level better than $10^{-6}$ between 2 $\lambda/D_L$ and 12 $\lambda/D_L$, and better than $3.10^{-7}$ between 7 $\lambda/D_L$ and 11 $\lambda/D_L$. The simulation performance was limited by the diffraction of the speckles of the uncorrected area in the focal plane created by the amplitude defects. In addition, in laboratory tests, the contrast is currently limited by the camera dynamics in the aberration estimation.

We experimentally proved that a small shrinking of the size of the correction zone can improve the contrast the contrast up to a factor 10. We analyzed the influence of the reference pupil radius on the performance of the SCC and proved that the reference of $\gamma = 16$ (the larger reference pupil possible with a nonzero reference flux inside the correction zone) provides the best correction in our case.

To enhance the performance of the self-coherent camera even more, we plan several improvements. First, one can directly minimize $A_S$, the speckle complex field measured by the SCC and not the phase estimated in the pupil plane. This approach has started to show good results \citep{BaudozSPIE12} for the simultaneous correction of amplitude and phase. The correction for the  amplitude errors can probably also be improved by the use of two DMs. Moreover, solutions are considered to use the SCC with wider spectral bandwidths. First tests in polychromatic light have already been conducted and show promising results~\citep{BaudozSPIE12}. A preliminary study of these effects has been published \citep{Galicher10}. A forthcoming paper will present a new version of the SCC that will probably overcome the current chromatic limitation.

\begin{acknowledgements}
J. Mazoyer is grateful to the Centre National d'Etudes Spatiales (CNES, Toulouse, France) and Astrium (Toulouse, France) for supporting his PhD fellowship. SCC development is supported by CNES (Toulouse, France).
 \end{acknowledgements}

\bibliographystyle{aa} 
\bibliography{bib_Maz2012}   

\begin{thebibliography}{31}
\expandafter\ifx\csname natexlab\endcsname\relax\def\natexlab#1{#1}\fi

\bibitem[{{Abe} {et~al.}(2003){Abe}, {Domiciano de Souza}, {Vakili}, \&
  {Gay}}]{Abe03}
{Abe}, L., {Domiciano de Souza}, Jr., A., {Vakili}, F., \& {Gay}, J. 2003,
  \aap, 400, 385

\bibitem[{{Baudoz} {et~al.}(2006){Baudoz}, {Boccaletti}, {Baudrand}, \&
  {Rouan}}]{Baudoz06}
{Baudoz}, P., {Boccaletti}, A., {Baudrand}, J., \& {Rouan}, D. 2006, in IAU
  Colloq. 2006: Direct Imaging of Exoplanets: Science \& Techniques, ed.
  {C.~Aime \& F.~Vakili}, 553--558

\bibitem[{{Baudoz} {et~al.}(2012){Baudoz}, {Mazoyer}, {Mas}, {Galicher}, \&
  {Rousset}}]{BaudozSPIE12}
{Baudoz}, P., {Mazoyer}, J., {Mas}, M., {Galicher}, R., \& {Rousset}, G. 2012,
  in Society of Photo-Optical Instrumentation Engineers (SPIE) Conference
  Series, Vol. 8446, Society of Photo-Optical Instrumentation Engineers (SPIE)
  Conference Series

\bibitem[{{Beuzit} {et~al.}(2008){Beuzit}, {Feldt}, {Dohlen}, {Mouillet},
  {Puget}, {Wildi}, {Abe}, {Antichi}, {Baruffolo}, {Baudoz}, {Boccaletti},
  {Carbillet}, {Charton}, {Claudi}, {Downing}, {Fabron}, {Feautrier},
  {Fedrigo}, {Fusco}, {Gach}, {Gratton}, {Henning}, {Hubin}, {Joos}, {Kasper},
  {Langlois}, {Lenzen}, {Moutou}, {Pavlov}, {Petit}, {Pragt}, {Rabou}, {Rigal},
  {Roelfsema}, {Rousset}, {Saisse}, {Schmid}, {Stadler}, {Thalmann}, {Turatto},
  {Udry}, {Vakili}, \& {Waters}}]{Beuzit08}
{Beuzit}, J.-L., {Feldt}, M., {Dohlen}, K., {et~al.} 2008, in Society of
  Photo-Optical Instrumentation Engineers (SPIE) Conference Series, Vol. 7014,
  Society of Photo-Optical Instrumentation Engineers (SPIE) Conference Series

\bibitem[{{Bord{\'e}} \& {Traub}(2006)}]{BordeTraub06}
{Bord{\'e}}, P.~J. \& {Traub}, W.~A. 2006, \apj, 638, 488

\bibitem[{{Boyer} {et~al.}(1990){Boyer}, {Michau}, \& {Rousset}}]{Boyer90}
{Boyer}, C., {Michau}, V., \& {Rousset}, G. 1990, in Society of Photo-Optical
  Instrumentation Engineers (SPIE) Conference Series, Vol. 1237, Society of
  Photo-Optical Instrumentation Engineers (SPIE) Conference Series, ed. J.~B.
  {Breckinridge}, 406--421

\bibitem[{{Cavarroc} {et~al.}(2006){Cavarroc}, {Boccaletti}, {Baudoz}, {Fusco},
  {Martinez}, \& {Rouan}}]{Cavarroc06}
{Cavarroc}, C., {Boccaletti}, A., {Baudoz}, P., {et~al.} 2006, in Society of
  Photo-Optical Instrumentation Engineers (SPIE) Conference Series, Vol. 6271,
  Society of Photo-Optical Instrumentation Engineers (SPIE) Conference Series

\bibitem[{{Galicher}(2009)}]{Galicher_these}
{Galicher}, R. 2009, PhD thesis, Universit\' e Denis Diderot Paris 7

\bibitem[{{Galicher} {et~al.}(2008){Galicher}, {Baudoz}, \&
  {Rousset}}]{Galicher08}
{Galicher}, R., {Baudoz}, P., \& {Rousset}, G. 2008, \aap, 488, L9

\bibitem[{{Galicher} {et~al.}(2010){Galicher}, {Baudoz}, {Rousset}, {Totems},
  \& {Mas}}]{Galicher10}
{Galicher}, R., {Baudoz}, P., {Rousset}, G., {Totems}, J., \& {Mas}, M. 2010,
  \aap, 509, A31+

\bibitem[{Give'on {et~al.}(2007)Give'on, Belikov, Shaklan, \&
  Kasdin}]{Giveon07}
Give'on, A., Belikov, R., Shaklan, S., \& Kasdin, J. 2007, Opt. Express, 15,
  12338

\bibitem[{{Give'on} {et~al.}(2006){Give'on}, {Kasdin}, {Vanderbei}, \&
  {Avitzour}}]{Giveon06}
{Give'on}, A., {Kasdin}, N.~J., {Vanderbei}, R.~J., \& {Avitzour}, Y. 2006, J.
  Opt. Soc. Am. A, 23, 1063

\bibitem[{{Guyon} {et~al.}(2009){Guyon}, {Matsuo}, \& {Angel}}]{guyonAPJ09}
{Guyon}, O., {Matsuo}, T., \& {Angel}, R. 2009, \apj, 693, 75

\bibitem[{{Guyon} {et~al.}(2005){Guyon}, {Pluzhnik}, {Galicher}, {Martinache},
  {Ridgway}, \& {Woodruff}}]{Guyon05}
{Guyon}, O., {Pluzhnik}, E.~A., {Galicher}, R., {et~al.} 2005, \apj, 622, 744

\bibitem[{{Kalas} {et~al.}(2008){Kalas}, {Graham}, {Chiang}, {Fitzgerald},
  {Clampin}, {Kite}, {Stapelfeldt}, {Marois}, \& {Krist}}]{Kalas08}
{Kalas}, P., {Graham}, J.~R., {Chiang}, E., {et~al.} 2008, Science, 322, 1345

\bibitem[{{Lagrange} {et~al.}(2009){Lagrange}, {Kasper}, {Boccaletti},
  {Chauvin}, {Gratadour}, {Fusco}, {Ehrenreich}, {Apai}, {Mouillet}, \&
  {Rouan}}]{Lagrange_Bpic09}
{Lagrange}, A., {Kasper}, M., {Boccaletti}, A., {et~al.} 2009, \aap, 506, 927

\bibitem[{{Macintosh} {et~al.}(2008){Macintosh}, {Graham}, {Palmer}, {Doyon},
  {Dunn}, {Gavel}, {Larkin}, {Oppenheimer}, {Saddlemyer}, {Sivaramakrishnan},
  {Wallace}, {Bauman}, {Erickson}, {Marois}, {Poyneer}, \&
  {Soummer}}]{Macintosh08}
{Macintosh}, B.~A., {Graham}, J.~R., {Palmer}, D.~W., {et~al.} 2008, in Society
  of Photo-Optical Instrumentation Engineers (SPIE) Conference Series, Vol.
  7015, Society of Photo-Optical Instrumentation Engineers (SPIE) Conference
  Series

\bibitem[{{Malbet} {et~al.}(1995){Malbet}, {Yu}, \& {Shao}}]{Malbet95}
{Malbet}, F., {Yu}, J.~W., \& {Shao}, M. 1995, \pasp, 107, 386

\bibitem[{{Marois} {et~al.}(2006){Marois}, {Lafreni{\`e}re}, {Doyon},
  {Macintosh}, \& {Nadeau}}]{Marois06a}
{Marois}, C., {Lafreni{\`e}re}, D., {Doyon}, R., {Macintosh}, B., \& {Nadeau},
  D. 2006, \apj, 641, 556

\bibitem[{{Marois} {et~al.}(2008){Marois}, {Macintosh}, {Barman}, {Zuckerman},
  {Song}, {Patience}, {Lafreni{\`e}re}, \& {Doyon}}]{Marois08}
{Marois}, C., {Macintosh}, B., {Barman}, T., {et~al.} 2008, Science, 322, 1348

\bibitem[{{Marois} {et~al.}(2004){Marois}, {Racine}, {Doyon}, {Lafreni{\`e}re},
  \& {Nadeau}}]{Marois04}
{Marois}, C., {Racine}, R., {Doyon}, R., {Lafreni{\`e}re}, D., \& {Nadeau}, D.
  2004, \apjl, 615, L61

\bibitem[{{Marois} {et~al.}(2010){Marois}, {Zuckerman}, {Konopacky},
  {Macintosh}, \& {Barman}}]{Marois10}
{Marois}, C., {Zuckerman}, B., {Konopacky}, Q.~M., {Macintosh}, B., \&
  {Barman}, T. 2010, Nature, 468, 1080

\bibitem[{{Mas} {et~al.}(2012){Mas}, {Baudoz}, {Rousset}, \&
  {Galicher}}]{Mas12}
{Mas}, M., {Baudoz}, P., {Rousset}, G., \& {Galicher}, R. 2012, \aap, 539, A126

\bibitem[{{Mas} {et~al.}(2010){Mas}, {Baudoz}, {Rousset}, {Galicher}, \&
  {Baudrand}}]{mmas_spie_2010}
{Mas}, M., {Baudoz}, P., {Rousset}, G., {Galicher}, R., \& {Baudrand}, J. 2010,
  in Society of Photo-Optical Instrumentation Engineers (SPIE) Conference
  Series, Vol. 7735, Society of Photo-Optical Instrumentation Engineers (SPIE)
  Conference Series

\bibitem[{{Mawet} {et~al.}(2005){Mawet}, {Riaud}, {Absil}, \&
  {Surdej}}]{Mawet05}
{Mawet}, D., {Riaud}, P., {Absil}, O., \& {Surdej}, J. 2005, \apj, 633, 1191

\bibitem[{{Mazoyer} {et~al.}(2012){Mazoyer}, {Baudoz}, {Mas}, {Rousset}, \&
  {Galicher}}]{MazoyerSPIE12}
{Mazoyer}, J., {Baudoz}, P., {Mas}, M., {Rousset}, G., \& {Galicher}, R. 2012,
  in Society of Photo-Optical Instrumentation Engineers (SPIE) Conference
  Series, Vol. 8442, Society of Photo-Optical Instrumentation Engineers (SPIE)
  Conference Series

\bibitem[{{Poyneer} \& {Macintosh}(2004)}]{Poyneer04}
{Poyneer}, L.~A. \& {Macintosh}, B. 2004, J. Opt. Soc. Am. A, 21, 810

\bibitem[{{Pueyo} {et~al.}(2010){Pueyo}, {Shaklan}, {Give'On}, {Troy},
  {Kasdin}, {Kay}, {Groff}, {McElwain}, \& {Soummer}}]{Pueyo10}
{Pueyo}, L., {Shaklan}, S.~B., {Give'On}, A., {et~al.} 2010, in Adaptative
  Optics for Extremely Large Telescopes

\bibitem[{{Rouan} {et~al.}(2000){Rouan}, {Riaud}, {Boccaletti}, {Cl{\'e}net},
  \& {Labeyrie}}]{Rouan00}
{Rouan}, D., {Riaud}, P., {Boccaletti}, A., {Cl{\'e}net}, Y., \& {Labeyrie}, A.
  2000, \pasp, 112, 1479

\bibitem[{{Trauger} \& {Traub}(2007)}]{Trauger07}
{Trauger}, J.~T. \& {Traub}, W.~A. 2007, \nat, 446, 771

\bibitem[{{Wallace} {et~al.}(2010){Wallace}, {Burruss}, {Bartos}, {Trinh},
  {Pueyo}, {Fregoso}, {Angione}, \& {Shelton}}]{Wallace10}
{Wallace}, J.~K., {Burruss}, R.~S., {Bartos}, R.~D., {et~al.} 2010, in Society
  of Photo-Optical Instrumentation Engineers (SPIE) Conference Series, Vol.
  7736, Society of Photo-Optical Instrumentation Engineers (SPIE) Conference
  Series

\end{thebibliography}

\end{document}